\documentclass[usegraphicx,usenatbib]{mn2e}
\begin{document}

\title[Stability of prograde and retrograde planets]{Stability of prograde and retrograde planets in circular binary systems}

\author[M.H.M. Morais and C.A. Giuppone]{M.H.M. Morais$^{1}$\thanks{helena.morais@ua.pt} and C.A. Giuppone$^{1}$ \\
$^{1}$Department of Physics \& I3N, University of Aveiro,
	     Campus Universit\'ario de Santiago, 3810-193 Aveiro, Portugal }
	     
\date{}

\maketitle

\begin{abstract}
We investigate the stability of prograde versus retrograde planets in circular binary systems  using numerical simulations. We show that retrograde planets are stable up to distances closer to the perturber than prograde planets. We develop an analytical model to compute the prograde and retrograde mean motion resonances'  locations and separatrices. We show that instability is due to  single resonance forcing, or caused by  nearby resonances' overlap.  
We validate our results regarding the role of single resonances and resonances' overlap on orbit stability, by computing surfaces of section of the CR3BP. We conclude that the observed enhanced stability of retrograde planets with respect to prograde planets is due to essential differences between the phase-space topology of retrograde versus prograde resonances (at $p/q$ mean motion ratio, prograde resonance is of order $p-q$ while retrograde resonance is of order $p+q$).
\end{abstract}

\section{Introduction}

The stability of coplanar prograde planet orbits in binary systems has been investigated numerically by \citet{Holman&Wiegert1999AJ}.
\citet{Mudryk&Wu2006ApJ} showed that instability in eccentric binaries is due to overlap of sub-resonances  associated with certain mean motion ratios $p/q$.  These sub-resonances are split due to the precession rate induced by the secondary star, hence  overlap of sub-resonances for a given mean motion ratio $p/q$ extends over a wide region and explains the instability regions in eccentric binaries.
Additionaly, \citet{Mudryk&Wu2006ApJ} suggest that the cause for instability in circular binaries is  overlap of sub-resonances  associated with the 3/1 mean motion ratio. However, this cannot work for circular binaries since in this case there is only one resonant angle.   

The circular restricted 3-body problem (CR3BP) is the simplest theoretical tool to understand planet stability within a binary system. 
The existence of an integral of the motion (Jacobi constant) reduces the number of variables of the coplanar problem from 4 to 3. Therefore, the phase space topology can be investigated by using surfaces of section for any given mass ratio $\mu$. The Jacobi constant and associated zero velocity curves (ZVC) can impose bounds on the test particle's motion. In particular, it is useful to compare the Jacobi constant  with the values at the collinear Lagrange points ($L_1$, $L_2$ and $L_3$).  When the Jacobi constant exceeds the value at $L_1$, the test particle must remain in orbit around either primary or secondary stars (concept of Hill stability, \citet{Szebehely1980}). When the Jacobi constant is smaller than the value at $L_1$, the test particle can orbit both stars and will eventually collide with one of them, although these capture episodes can be long-lived \citep{Winter&Neto2001,Astakhov_etal2003}. When the Jacobi constant is smaller than the values at $L_2$ or $L_3$, the test particle can escape through these points. However, its is well known that these are  necessary but not sufficient conditions for instability  \citep{Szebehely1980}.  

\citet{Eberle_etal2008A&A} investigated stability of prograde planet orbits within circular binary systems, based on the Jacobi constant criterion. \citet{Quarles_etal2011A&A} validated these results by computing  the maximum Lyapunov exponent which is a measure of chaos and associated instability.
\citet{Quarles_etal2011A&A} show that if the ZVC opens at L3 then the orbit is unstable but when the ZVC opens at L1 or L2, the orbit is not necessarily unstable. The interpretation of these results is not obvious and it may depend on the particular choice of initial conditions.

\citet{Chirikov1960} established the resonance overlap  criterion to explain the onset of chaotic motion in Hamiltonian deterministic systems.
\citet{Wisdom1980AJ} obtained a resonance overlap criterion for the onset of chaos in the CR3BP valid when 
$\mu\ll 1$. \citet{Wisdom1980AJ}  showed that first order mean motion resonances with the secondary overlap in a region of width $\sim\mu^{2/7}$.  Orbits in this region exhibit chaotic diffusion of eccentricity and semi-major axis until escape or collision occurs.    
When $\mu\ll 1$ individual resonances cannot increase the eccentricity or semi-major axis up to escape values.
However, in binary star systems $\mu\sim 1$ thus the effect of single resonances is not necessarily negligible.

Recently, it was possible to detect  the Rossiter-MacLaughin effect on transiting extra-solar planets \citep{Triaud2010}. This effect allows to measure  the orientation of the planet' orbit with respect to the parent star's equator. Contrary to what happens in the Solar System, extra-solar planets' orbits can be misaligned with the host star's equator with angles that range from $0^\circ$ to $180^\circ$.  
Several mechanism have been proposed to explain these misaligned planets. These include classic Kozai oscillations due to a nearby star with subsequent tidal drift \citep{Correia_etal2011CMDA},  secular interaction with a companion brown dwarf or giant planet followed by tidal drift \citep{Naoz_etal2011Nature},  secular chaos and tides in multiple planet systems \citep{Wu&Lithwick2011}, orbit interaction in multiple planet systems with a companion star \citep{Kaib_etal2011ApJ}, or planet-planet scattering and tides \citep{Beauge&Nesvorny2012}.

\citet{Gayon&Bois2008} used the MEGNO chaos indicator \citep{Cincotta_2000} to show that retrograde resonance in 2 planet systems is more stable than the equivalent prograde resonance.  
Therefore, they suggest that retrograde resonance could explain the radial velocity  data of extra-solar systems where prograde resonance is unstable \citep{Gayon&Bois2009MNRAS}.  In \citet{Gayon_etal2009CMDA}, an expansion of the Hamiltonian for retrograde resonance in 2 planet systems is developed. However, the numerical exploration in \citet{Gayon_etal2009CMDA} is limited to a small set of initial conditions which could explain why they do not conclude  on the essential differences between prograde and retrograde resonance. 

Planets in retrograde orbits  within a binary system are a theoretical possibility although none was confirmed to date.
A planet on a retrograde orbit has been suggested as an explanation for a periodic signal  of 416~d in  $\nu$-Octantis A radial velocity curve \citep{Ramm_etal2009MNRAS}. The star  $\nu$-Octantis A  orbits its companion ($\nu$-Octantis B) on a 2.9~yrs  orbit. Such a tight binary orbit implies that a prograde planet  at 416~d is unstable but a retrograde planet could be stable at least up to $10^6$~yrs \citep{Eberle&Cuntz2010ApJL}. Nevertheless, there are alternative hypothesis that claim that  $\nu$-Octantis A radial velocity could be explained, without the need of a planet, if $\nu$-Octantis B was a double star \citep{Morais&Correia2012MNRAS}.

The purpose of this article is to  investigate stability of coplanar prograde and retrograde planet orbits in circular binary systems. Contrary to previous works, we will  not only perform  simulations but  will also provide  theoretical explanations for the onset of instability based on the effect of single resonances or due to  resonance overlap.

\section{Expansion of the disturbing function in CR3BP}

We consider the planar CR3BP composed of a test particle orbiting a primary $m_0$, and perturbed by a secondary $m_2$. The primary and secondary have  a circular orbit with frequency $n_2=\sqrt{G (m_0+m_2)/a_2^3}$ and radius $a_2$. 
Since we want to model the perturbation from the secondary we write the equation of motion in the frame with origin at the primary 
\begin{equation}
{\bf \ddot{r}}_1=-\nabla \left( U_0 + U \right) \ ,
\end{equation}
where $U_0=G\,m_0/r_1$, $G$ is the gravitational constant, and $U$ is the disturbing potential due to the perturber $m_2$.

When $U=0$ (i.e.\ $m_2=0$ ) the solution to Eq.~(1) is a Keplerian elliptical orbit with mean motion $n_1=\sqrt{G\,m_0/a_1^3}$, semi-major axis  $a_1$, eccentricity $e_1$, longitude of  pericenter $\varpi_1$, and  true anomaly $f_1$.

The disturbing potential is
\begin{equation}
\label{disturbingf}
U= G\,m_2 \left( \frac{1}{\Delta} - \frac{\alpha}{a_{2}}\frac{r_1}{a_1}\,\cos{S} \right) \ ,
\end{equation} 
where
\begin{equation}
\Delta=||{\bf r}_1-{\bf r}_2||={\sqrt{r_1^2+a_2^2-2\,r_1\,a_2\,\cos{S}}} \ ,
\end{equation}
$\alpha=a_1/a_2<1$, and  $S$ is the angle between ${\bf r}_1$ and ${\bf r}_2$.

 In the prograde case the primary-secondary and test particle orbit in the same direction. In the retrograde case the test particle orbits in the opposite direction of the primary-secondary. The primary-secondary relative position vector is ${\bf r}_2=a_2\,(\cos(\lambda_2),\sin(\lambda_2))$.The test particle ($m_1=0$) position vector with respect to the primary is ${\bf r}_1=r_1\,(\cos(f_1+\varpi_1),\sin(f_1+\varpi_1))$ with $r_1=a_1\,(1-e_1^2)/(1+e_1\,\cos{f_1})$.
Hence,
\begin{equation}
\cos{S} = \cos(f_1+\varpi_1-\lambda_2) \ ,
\end{equation}
where $\lambda_2=\pm n_2\,t$, and the $\pm$ sign applies to the prograde or retrograde cases, respectively.

The 1st and  2nd terms in  Eq.~(2) are known as direct and indirect parts, respectively.  The disturbing potential (Eq.~(2) can be expressed in the orbital elements $(a_1,e_1,f_1,\varpi_1)$ by using Laplace coefficients (literal expansion). The direct part is written as a Taylor series in $\epsilon=(r_1/a_1-1)$ i.e.
\begin{equation}
\frac{1}{\Delta}=\left( 1+\sum_{i=1}^{\infty}\frac{1}{i!}\,\epsilon^i\,\alpha^i\,\frac{d^i}{d_i \alpha} \right) \frac{1}{\rho}
\end{equation}
with
\begin{eqnarray}
\frac{1}{\rho}&=&\frac{1}{a_2} (1+\alpha^2-2\,\alpha\,\cos{S})^{-1/2} \nonumber \\
 &=& \frac{1}{a_2} \sum_{j} \frac{1}{2} b_{1/2}^{j}(\alpha) \cos(j\,S) 
\end{eqnarray} 
where $b_{1/2}^{j}(\alpha)$ is a Laplace coefficient.

Since
\begin{eqnarray}
\cos(j\,S) &=& \cos(j (f_1+\varpi_1- \lambda_2)) \  ,
\end{eqnarray}
using elliptic expansions for $r_1/a_1$, $\cos{f_1}$ and $\sin{f_1}$, we obtain for any given $j$,  the direct and indirect parts of Eq.~(\ref{disturbingf}).  This is done in \citet{Ellis&Murray2000} for prograde resonances. In the planar CR3BP, we only consider those terms in the expansion from \citet{Ellis&Murray2000} that depend on $e_1$. These terms consist of  cosines of angles which are combinations of the mean longitudes $\lambda_2=n_2\,t$,
$\lambda_1=n_1\,(t-\tau)+\varpi_1$ (where $\tau$ is the time of passage at pericenter), and the longitude of pericenter $\varpi_1$.  From the discussion above we conclude that for retrograde resonances the terms are exactly the same, although we must replace $\lambda_2=-n_2\,t$.

By inspecting the expansion of the disturbing potential in \citet{Ellis&Murray2000} we see that at 1st order in $e_1$, terms of the type $(j-1)\,\lambda_1-j\,\lambda_2+\varpi_1$ appear (4D1.1 in \citet{Ellis&Murray2000}).  If $j\ge 2$, these terms correspond to a $j/(j-1)$ prograde resonance since $\dot{\lambda}_1=n_1$ and $\dot{\lambda}_2=n_2$ thus the time variation of the angle is $(j-1)\,n_1-j\,n_2 \approx 0$. In the retrograde case, $\dot{\lambda}_2=-n_2$ hence the previous terms are non-resonant.  At 2nd order in $e_1$ terms of the type $(j-2)\,\lambda_1-j\,\lambda_2+2\,\varpi_1$  appear (4D2.1 in \citet{Ellis&Murray2000}) which correspond to a $j/(j-2)$ prograde resonance ($j\ge 3$) or the $1/1$ retrograde resonance when $j=1$. At 3rd order in $e_1$ terms of the type $(j-3)\,\lambda_1-j\,\lambda_2+3\,\varpi_1$ appear (4D3.1 in \citet{Ellis&Murray2000}) which correspond to a $j/(j-3)$ prograde resonance ($j\ge 4$) or the $2/1$ retrograde  resonance when $j=2$.
 At 4th order in $e_1$ terms of the type $(j-4)\,\lambda_1-j\,\lambda_2+4\,\varpi_1$ appear (4D4.1 in \citet{Ellis&Murray2000}) which correspond to a $j/(j-4)$ prograde resonance ($j\ge 5$) or the $3/1$ retrograde resonance when $j=3$.
It can be shown that at 5th order in $e_1$ terms of the type $(j-5)\,\lambda_1-j\,\lambda_2+5\,\varpi_1$ appear which correspond to a $j/(j-5)$ prograde resonance ($j\ge 6$) or to the $3/2$ retrograde resonance when $j=3$ and the $4/1$ retrograde resonance when $j=4$.
Therefore, we see that $p/q$ prograde resonances are of order $p-q$ while $p/q$ retrograde resonances are of order $p+q$. The only resonant terms in the indirect part of the disturbing function  (CR3BP) correspond to the $1/1$ prograde resonance (4E0.1 in \citet{Ellis&Murray2000}) and to the $1/1$ retrograde resonance (4E2.2 in \citet{Ellis&Murray2000}). The secular term is obtained by averaging over the mean longitudes $\lambda_1$ and $\lambda_2$, hence it is the same in the prograde or retrograde case (4D0.1 in \citet{Ellis&Murray2000}).

\section{Analytic model for  mean motion resonance in CR3BP}

Here, we briefly review the analytic model for 1st, 2nd and 3rd order prograde resonance from \citet{ssdbook}. In Appendix A  we derive in detail this Hamiltonian model  in the framework of the CR3BP. We extend the model to retrograde resonance of lowest (3rd) order. We  explain how we can use the model to obtain resonance widths for initially circular orbits.

\subsection{Hamiltonian model for prograde/retrograde resonance}

In the CR3BP, $j/(j-k)$ prograde resonance is of order $k$ while $j/(k-j)$ retrograde resonance\footnote{We will use the notation $j/(k-j)$ retrograde resonance or $j/(j-k)$ resonance: e.g.~ 2/1 retrograde resonance or 2/-1 resonance ($j=2$, $k=3$).}
is of order $k$.
The resonant angle is
\begin{equation}
\theta= (j-k)\,\lambda_1- j\,\lambda_2 + k\,\varpi_1 \ .
\end{equation}

Here, we will summarize the results for prograde  $j/(j-k)$ resonances of 1st, 2nd and 3rd order ($k=1,2,3$)
and we will extend these results to the retrograde $2/1$ resonance which is of 3rd (lowest) order ($j=2$, $k=3$).
The resonant Hamiltonian (Eq.~\ref{hamilton0}) depends on a single parameter (Eq.~\ref{delta})
\begin{equation}
\label{delta0}
 \delta = A [ (j-k)\,n_{1}^{*} \mp j\,n_2 + k\,\dot{\varpi}^{*}_{1} ]/k \ ,
 \end{equation}
where $n_{1}^{*}=n_1+\dot{\lambda}^{*}_{1}$,  
\begin{equation}
\label{scalingdelta}
A= \left( \frac{2^{4-k}}{3^{2-k}}  \frac{ j^{k-8/3} (j-k)^{k-4/3} k^{4-2\,k}}{\mu^2\,f_d(\alpha)^2} \right)^{\frac{1}{4-k}} \ ,
\end{equation}
and from Eqs.~(\ref{sec1},\ref{sec2}) with $e_1\ll 1$
\begin{eqnarray}
\label{precession}
\dot{\varpi}^{*}_{1} &\approx & 2\,\frac{m_2}{m_0}\,\alpha\,f_{s}(\alpha)\,n_1 \\
\dot{\lambda}^{*}_{1} &\approx & \frac{e_1^2}{2}  \dot{\varpi}^{*}_{1} 
\end{eqnarray}
with values of $f_{s}(\alpha)$ at resonant  $\alpha=[j/|j-k|]^{-2/3}$ shown in Table~1.

The Hamiltonian  (Eq.~\ref{hamilton0}) is expressed in  cartesian canonical variables
\begin{eqnarray}
x &=& R\,\cos(\theta/k) \\
y &=&  R\,\sin(\theta/k)
\end{eqnarray}
where the scaling factor is  (Eq.~\ref{scaling0})
\begin{equation}
\label{scaling00}
R=\left[ \frac{3\,(-1)^k}{f_d(\alpha)\,\mu} \frac{(j-k)^{4/3} j^{2/3}}{k^{2}} \right]^\frac{1}{4-k}\,e_1 \  
\end{equation}
with values of $f_d(\alpha)$ at resonant  $\alpha=[j/|j-k|]^{-2/3}$ shown in Table~1.

\begin{table}
\centering
\begin{tabular}{|c||c|c|c|c|}
\hline
   resonance &  $\alpha$ & $\alpha\,f_{s}(\alpha)$  & $\alpha\,f_d(\alpha)$ \\
\hline
4/1 & 0.39685 & 0.032355  &  -0.09698\\
3/1 & 0.48075 & 0.068381   &  +0.28785 \\
5/2 & 0.54288 & 0.11600  &  -0.61503\\
2/1 & 0.62996  & 0.24419   & -0.74996 \\
2/-1 & 0.62996  & 0.24419   &  -0.25304 \\
5/3 &  0.71138 & 0.51566  &   +2.32892 \\
3/2 & 0.76314  & 0.87975   &  -1.54553 \\
\hline
\end{tabular}
\caption{Values of secular an resonant functions at resonant $\alpha=(j/|j-k|)^{-2/3}$, $k=1,2,3$.}
\end{table}

Obviously, this analytic model is valid  only for small perturbation i.e.\ (Eqs.~\ref{h0},\ref{resonant},\ref{secular})
\begin{eqnarray}
\label{perturbation}
\frac{U_{res}}{H_0} &=& \frac{1}{2}\frac{m_2}{m_0}\,\alpha\,f_d(\alpha)\,e_1^{k} \ll 1 \ , \\
\label{perturbationsec}
\frac{U_{sec}}{H_0} &=& \frac{1}{2}\frac{m_2}{m_0}\,\alpha\,f_s(\alpha)\,e_1^{2} \ll 1 \ .
\end{eqnarray}
In Table~1 we show the values of $\alpha\,f_d(\alpha)$ and $\alpha\,f_s(\alpha)$ at resonant value $\alpha=[j/|j-k|]^{-2/3}$. These provide a measure of the analytic model validity. In particular, the resonance location ($\delta=0$, Eq.~\ref{delta0}) depends on the secular term ($\dot{\varpi}_1$, Eq.~\ref{precession}), hence it is only accurate when Eq.~(\ref{perturbationsec}) is verified. From Table.~1 we see that the secular term ( Eq.~(\ref{perturbationsec})) at the 4/1 resonance is about 0.47, 0.28 and 0.13 times that at the 3/1, 5/2 and 2/1 (or 2/-1) resonances, respectively. Therefore, the secular term (Eq.~(\ref{perturbationsec})) is approximately the same when $\mu=0.2$, $\mu=0.09$, $\mu=0.07$ and $\mu=0.03$ at the 4/1, 3/1, 5/2 and 2/1 (or 2/-1) resonances, respectively. 

\subsection{First, second and third order resonances}

\begin{figure}
  \centering
  \includegraphics[width=7.5cm]{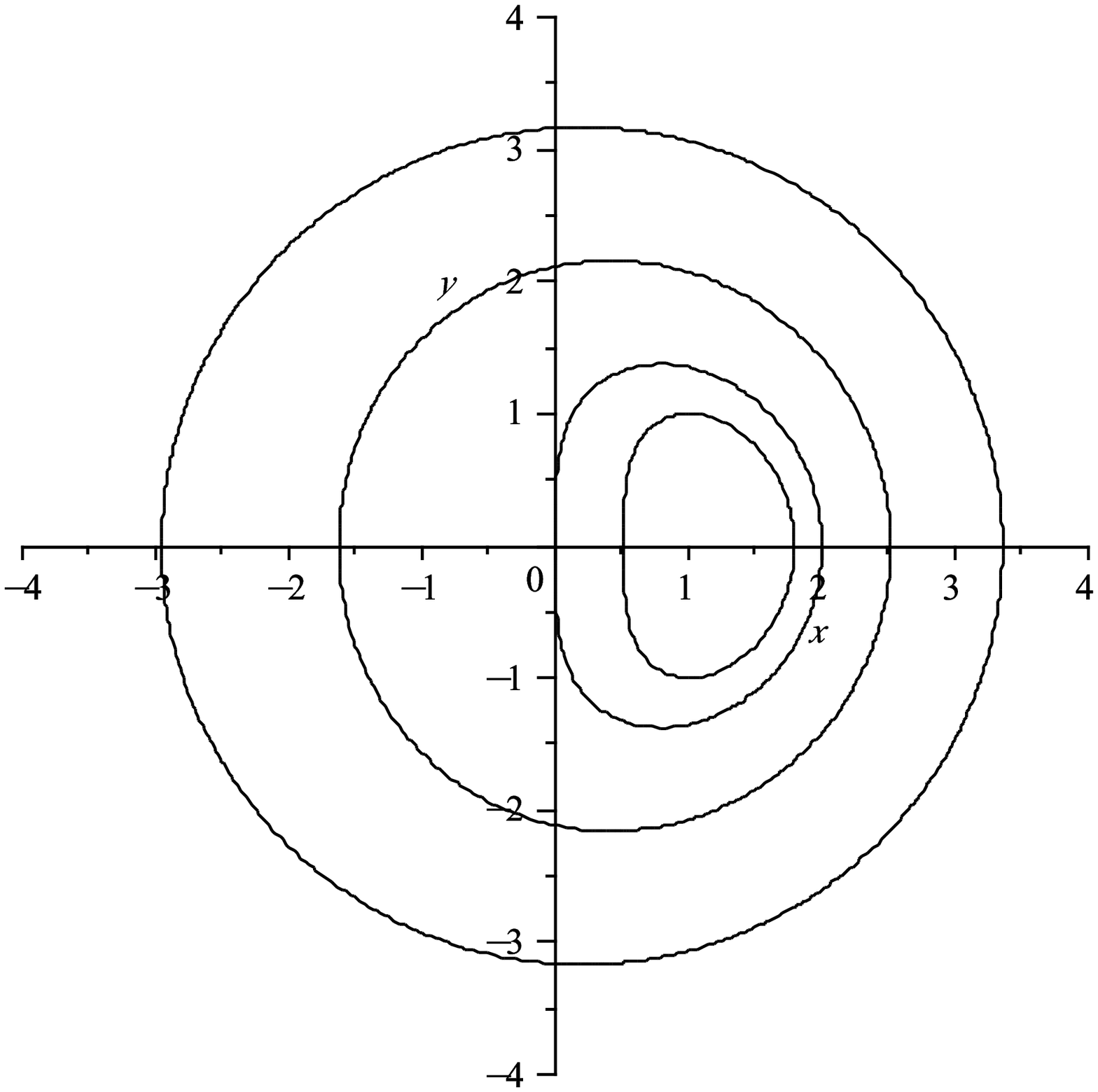} (a)
  \includegraphics[width=7.5cm]{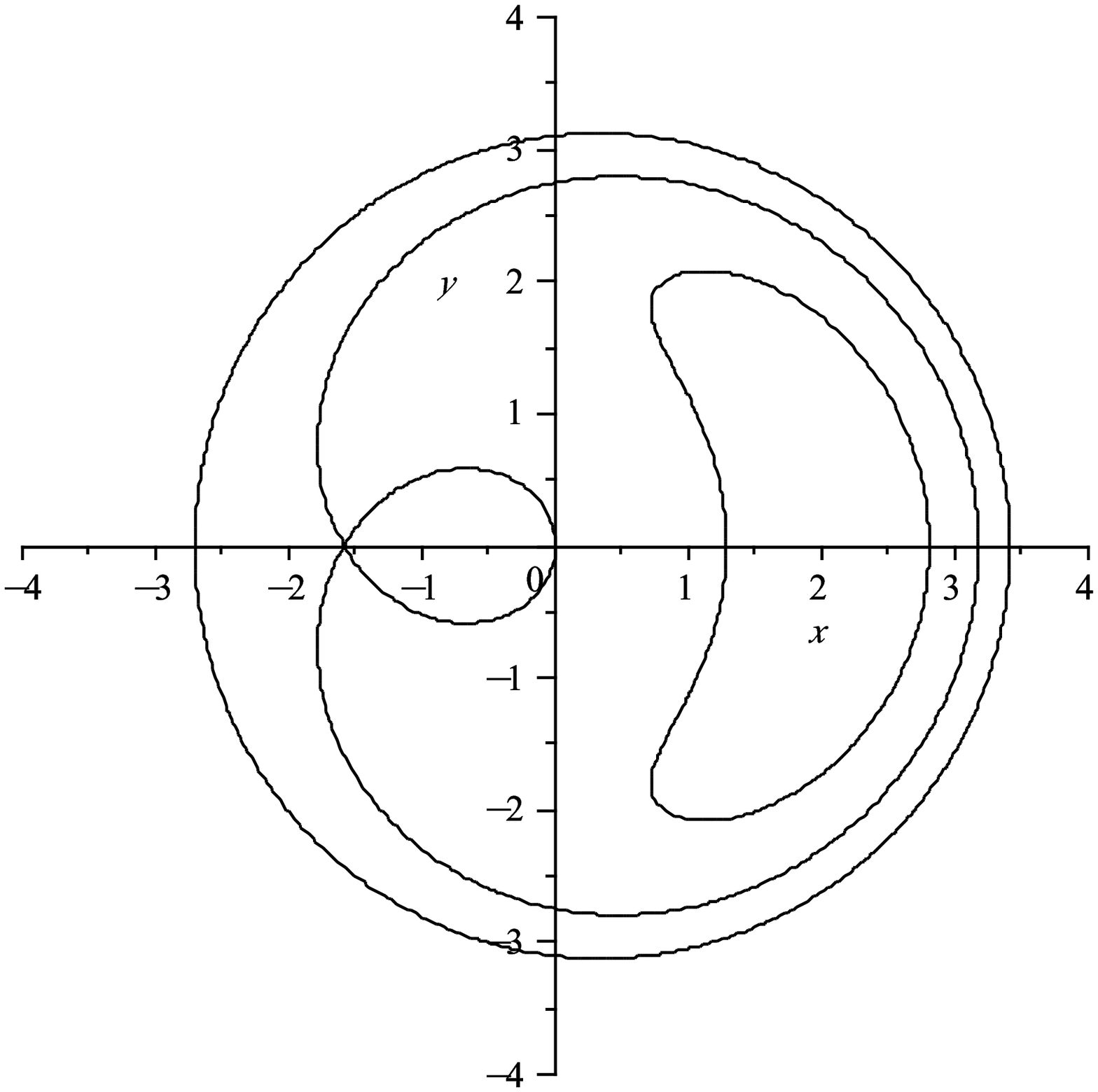}  (b)
\caption{Curves of constant $H_1$: $\delta=0$ (a) and $\delta=-3.78$ (b).  The separatrix intersects the origin at $\delta=-3.78$.}
\label{h1}
\end{figure}

The Hamiltonian (Eq.~\ref{hamilton0}) when $k=1$ is 
\begin{equation}
H_1=\frac{\delta}{2}\,(x^2+y^2)+\frac{1}{4}\,(x^2+y^2)^2-2\,x \ . 
\end{equation}
When $\delta<-3$ there is a single stable equilibrium point (Fig.~\ref{h1}a). At $\delta=-3$ an unstable equilibrium point appears which bifurcates into a stable/unstable pair visible when  $\delta<-3$ (Fig.~\ref{h1}b). 

\begin{figure}
  \centering
  \includegraphics[width=7.5cm]{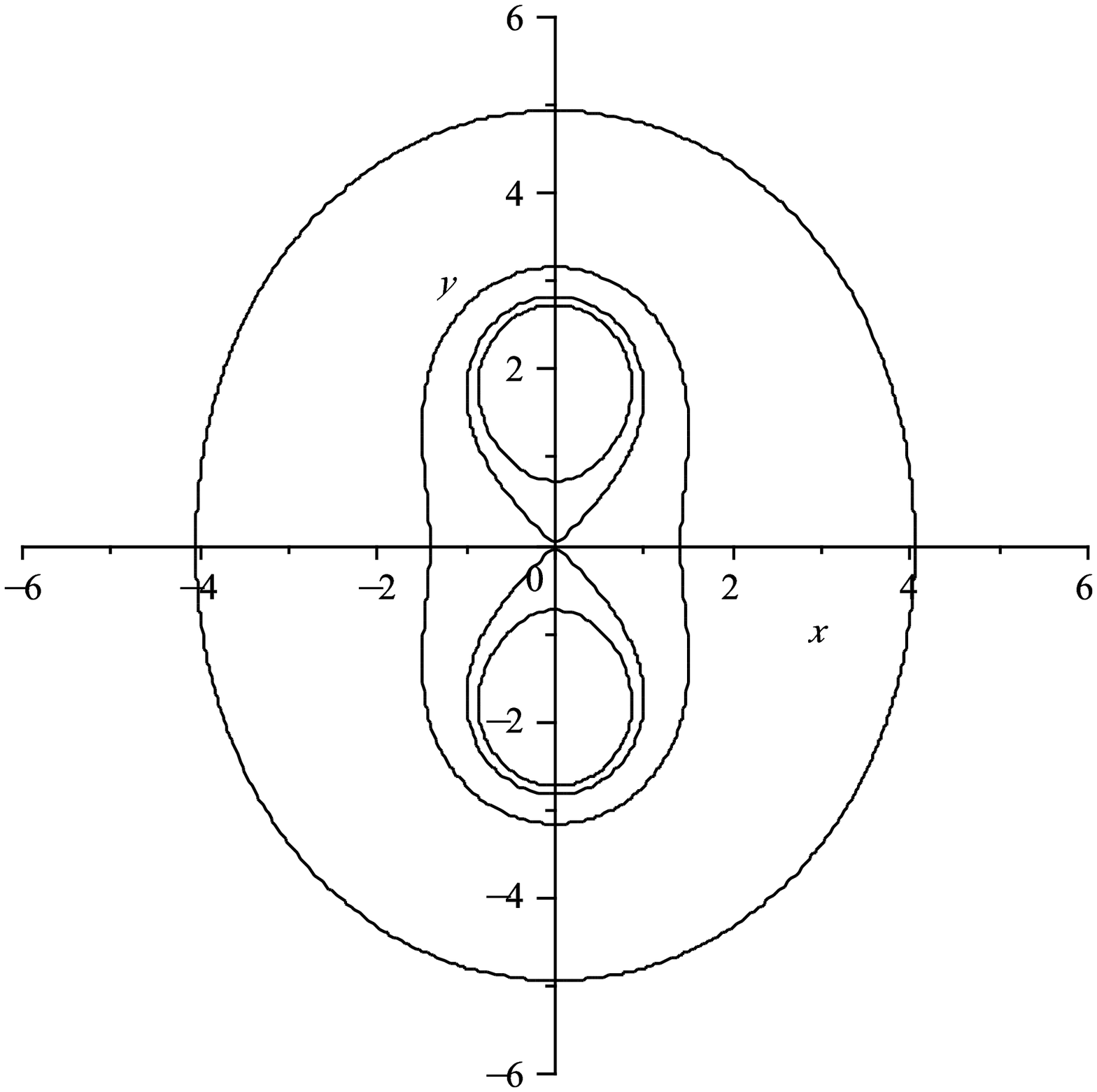} (a)
  \includegraphics[width=7.5cm]{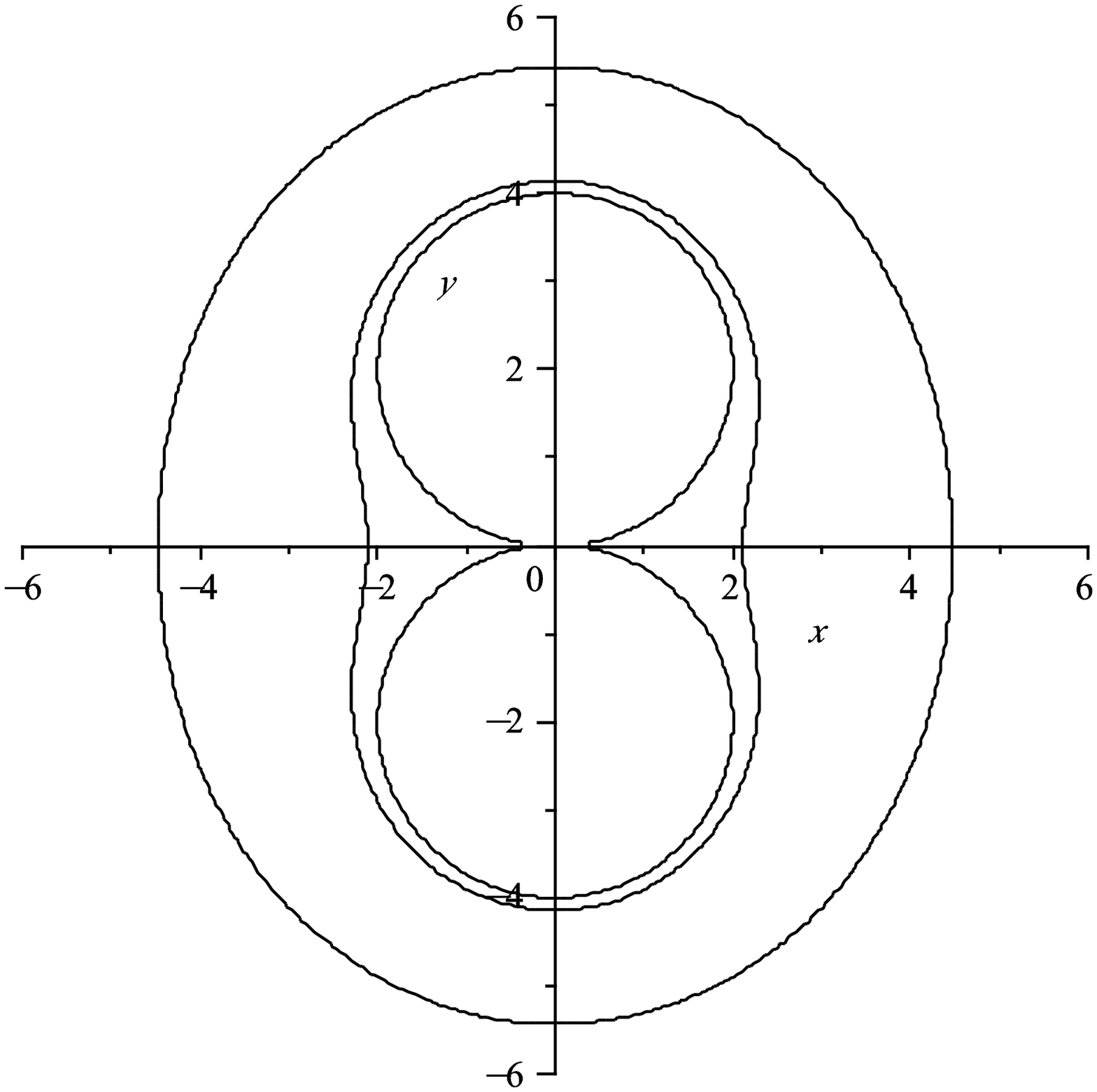}  (b)
  \includegraphics[width=7.5cm]{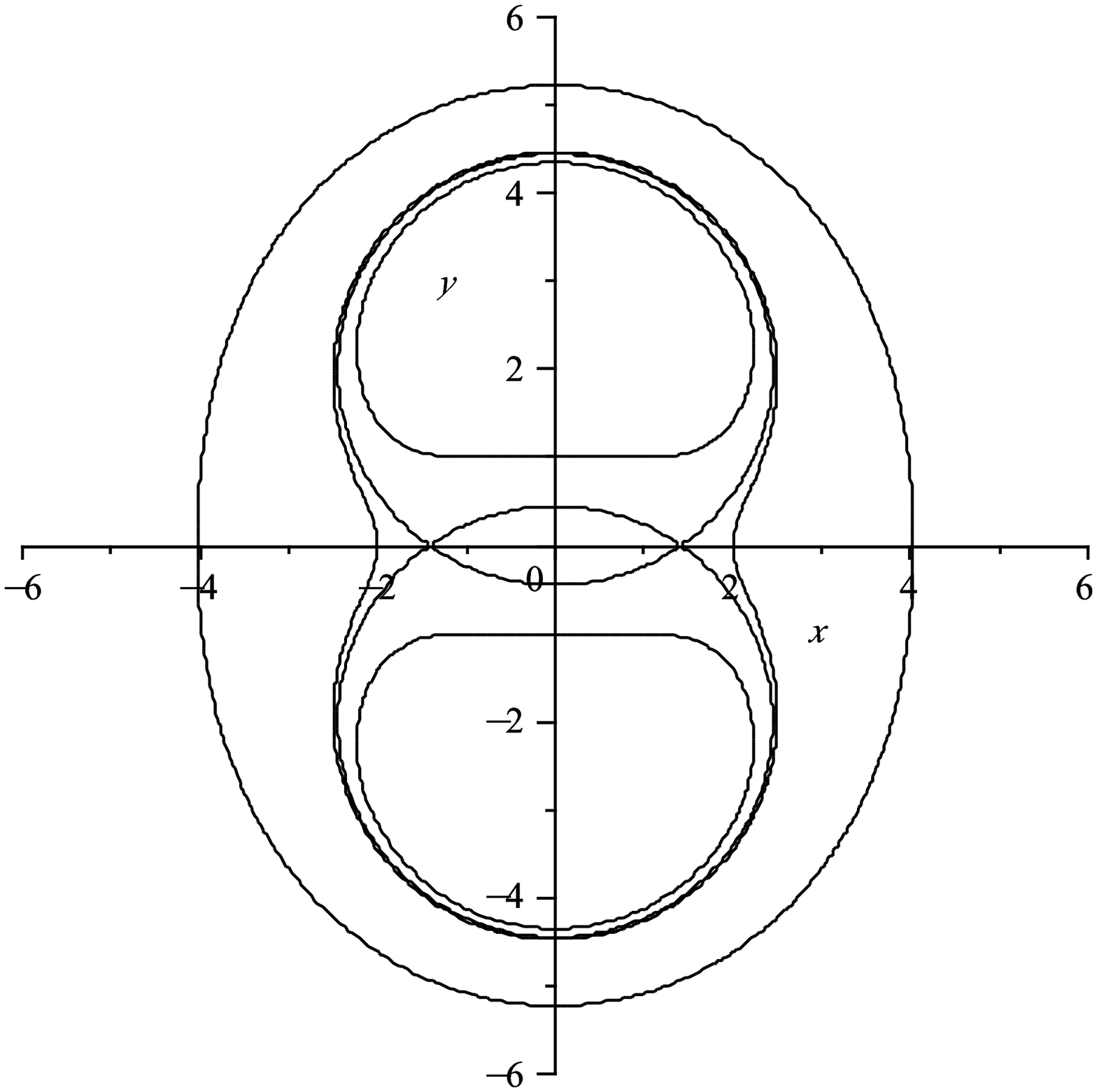}  (c)
\caption{Curves of constant $H_2$: $\delta=0$ (a), $\delta=-4$ (b) and $\delta=-6$ (c).  The separatrix intersects the origin between $\delta=0$ and $\delta=-4$.}
\label{h2}
\end{figure}

The Hamiltonian (Eq.~\ref{hamilton0}) when $k=2$ is 
\begin{equation}
H_2=\frac{\delta}{2}\,(x^2+y^2)+\frac{1}{4}\,(x^2+y^2)^2+2\,(x^2-y^2) \ . 
\end{equation}
At $\delta=4$ the origin becomes an unstable point and 2 stable points appear at $\phi=\pm\pi/2$ which move away from the origin as $\delta$ increases (see $\delta=0$ (Fig.~\ref{h2}a) and $\delta=-4$ (Fig.~\ref{h2}b)).
At $\delta=-4$ the origin becomes again a stable point (Fig.~\ref{h2}b)  and 2 unstable points appear at $\phi=0,\pi$ that are visible when $\delta<-4$ (Fig.~\ref{h2}c).

\begin{figure}
  \centering
   \includegraphics[width=7.5cm]{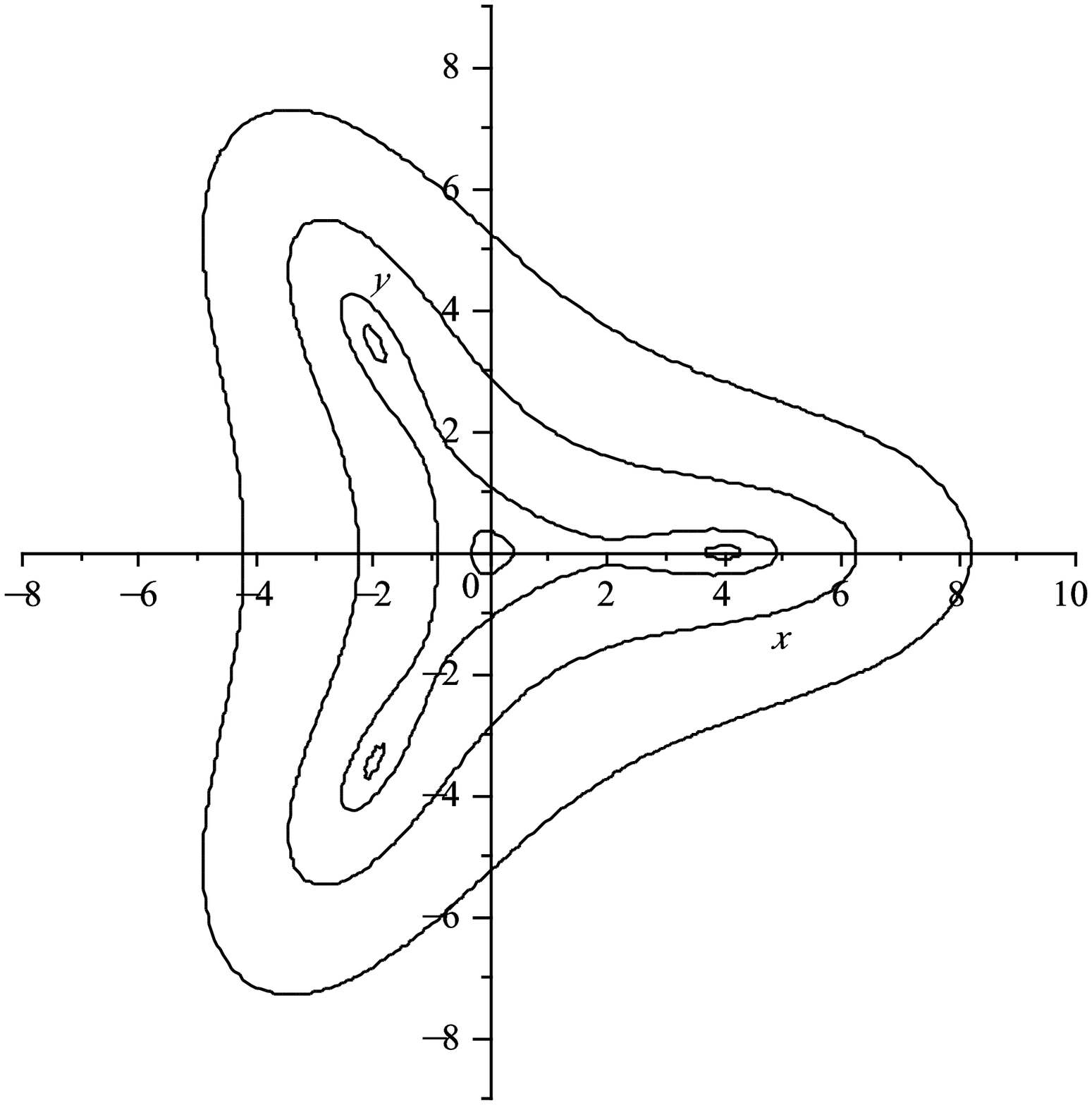} (a)
  \includegraphics[width=7.5cm]{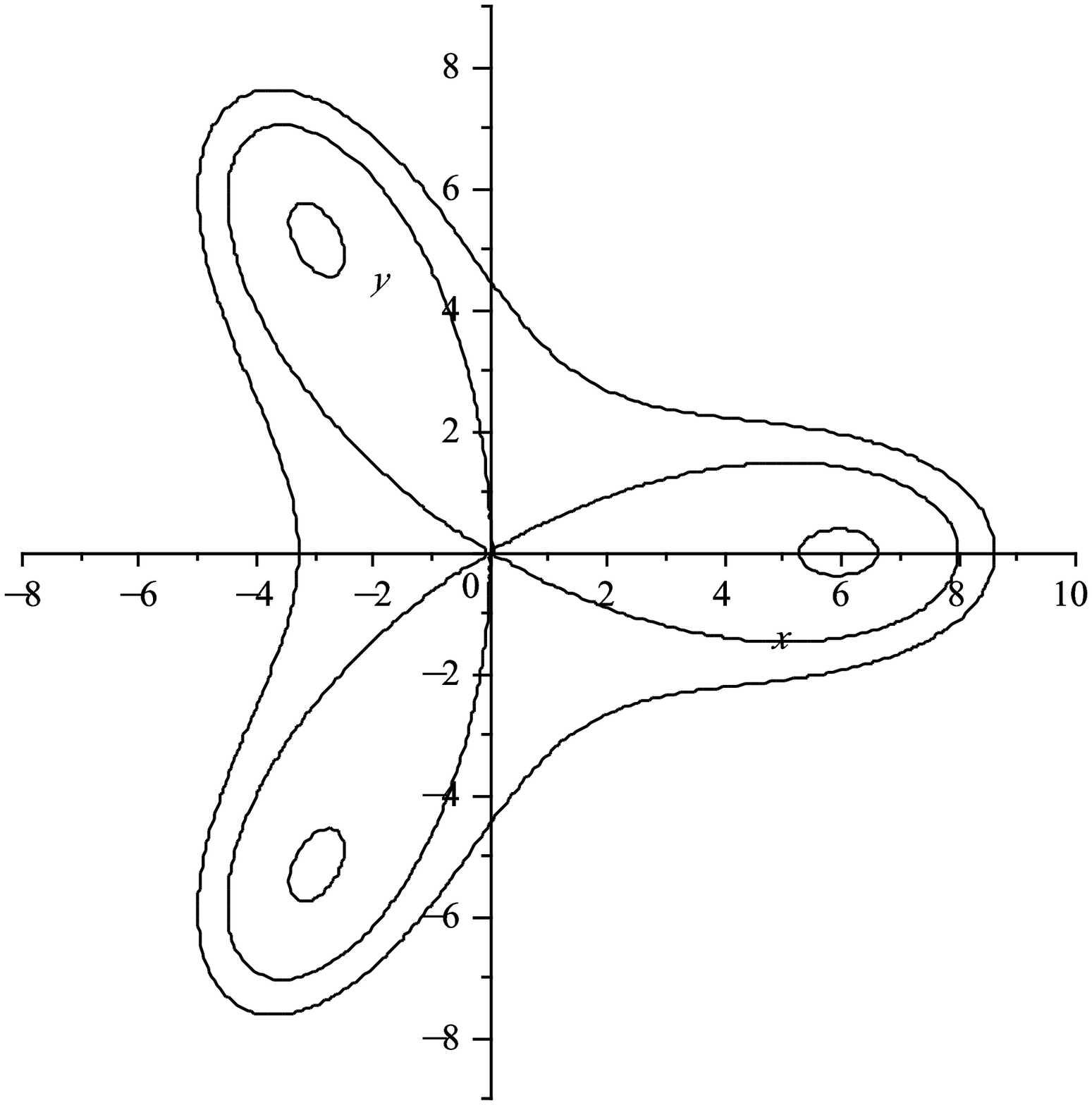}  (b)
   \includegraphics[width=7.5cm]{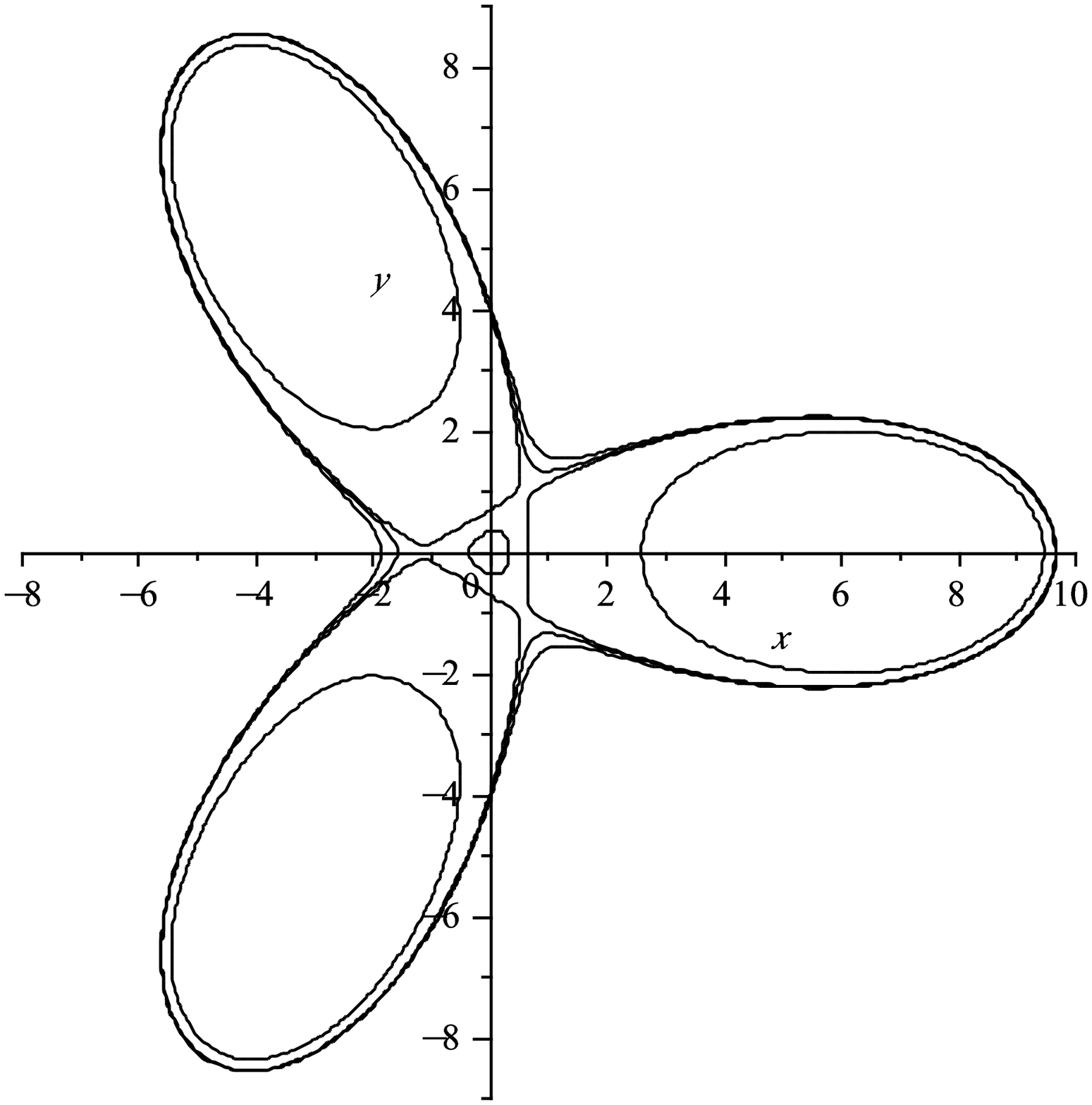}  (c)
\caption{Curves of constant $H_3$: $\delta=8$ (a), $\delta=0$ (b) and $\delta=-8$ (c).  The separatrix intersects the origin only at $\delta=0$ (exact resonance).}  
\label{h3}
\end{figure}

The Hamiltonian (Eq.~\ref{hamilton0}) when $k=3$ is\footnote{The diagrams with curves of constant $H_3$ in \citet{ssdbook} should be rotated by $\pi/3$.} 
\begin{equation}
H_3=\frac{\delta}{2}\,(x^2+y^2)+\frac{1}{4}\,(x^2+y^2)^2-2\,x\,(x^2-3\,y^2) \ . 
\end{equation}
At $\delta=9$ the origin is a stable point and 3 pairs of stable/unstable points appear at $\phi=0,\pm 2\,\pi/3$. The 3 stable points move away from the origin while the 3 unstable points move towards the origin (Fig.~\ref{h3}a), until they coincide with it at $\delta=0$ (Fig.~\ref{h3}b).   At  $\delta=0$ (exact resonance) the origin bifurcates into a stable point  and 3 unstable points  at $\phi=\pm\pi/3,\pi$. These unstable points move away from the origin as $\delta$ decreases (Fig.~\ref{h3}c).

\subsection{Computing the  resonances' separatrices}

In order to compute the resonances' widths we follow the method described in \citet{Wisdom1980AJ} for 1st order mean motion resonances.  This method was developed specifically for initially circular orbits and does not rely on the pendulum approximation which is more appropriate near the resonance center at $e_1\neq0$. Wisdom's method consists in measuring the variation in the parameter, i.e.\ $\Delta\delta$, between exact resonance ($\delta=0$) and the last value at which the separatrix intersects the origin (i.e.\  when the orbit with $e_1=0$ is at the separatrix). From Eq.~(\ref{delta0}), we have
\begin{equation}
 |\Delta\delta|=A\,\frac{|j-k|}{k}\,\Delta\,n_1
 \end{equation}
and using Kepler's 3rd law we obtain the resonance half-width 
\begin{equation}
\label{wisdom}
\frac{\Delta\,a_1}{a_1}=\frac{2\,k}{3\,j}\frac{|\Delta\delta|}{A} \ .
\end{equation}

 For 1st order resonances the separatix intersects the origin at $\delta\approx -4$ (Fig.~\ref{h1}(b)) while for 2nd order resonances the separatrix intersects the origin from $\delta=0$ (Fig.~\ref{h2}(a)) to $\delta=-4$ (Fig.~\ref{h2}(b)). Hence, we take a range $|\Delta\delta| \approx 4$ and apply Eq.~\ref{wisdom} to obtain the resonances' half-widths.
We checked that the 1st order resonances' widths are in agreement with Wisdom's approximate expressions \citep{Wisdom1980AJ}.

For 3rd order resonances the separatrix only intersects the origin at $\delta=0$ (Fig.~\ref{h3}(b)). Hence, $\Delta\delta=0$ i.e.\ 3rd order resonances have zero width. This  predicted zero width  at $e_1=0$ is a known feature of the analytic models,  e.g.\ it also occurs when using the pendulum approximation \citet{ssdbook,Mudryk&Wu2006ApJ}.

The exact resonance location  is obtained by solving $\delta=0$  (Eq.~\ref{delta0}) for $\alpha=a_1/a_2$.
In Sect.~5 we will see that, at small to moderate $\mu$ values, the resonance's widths / locations are in  reasonably good agreement with the numerical results obtained by the method of surfaces of section.

 \section{Initial conditions and zero velocity curves}

We chose a binary system with masses $m_0=M_{\odot}$ (primary) and $m_2\le M_{\odot}$ (secondary), inter-binary distance $a_2=1$~AU, and mass ratio $\mu=m_2/(m_0+m_2)$.
We chose units such that $G\,(m_0+m_2)=1$ which implies a binary period $T_2=2\,\pi$~yrs. The initial orbital elements with respect to the primary were semi-major axis $a_1$, eccentricity $e_1=0$,  mean longitude $\lambda_1=0$, inclination $I=0$ (prograde orbits) or $I=\pi$ (retrograde orbits). Hence,  the test particle was  always started between the primary and secondary, orbiting in the same direction (prograde orbit) or in the opposite direction (retrograde orbit). 

The planet's orbit can remain bounded to the primary (stable orbit), or it can become unstable. Unstable orbits collide with either primary or secondary, or escape from the system.
We assume collision with the primary if $r_0<0.005$~AU (i.e.\ the test particle gets within one solar radius of $m_0$), collision with the secondary if $r_0<0.005\,m_2/m_0$~AU, and escape from the system if $r>3$~AU. In practice, temporary capture in chaotic orbits around the secondary is possible but all  these capture episodes end either by  collision with the stars or escape from the system.

The CR3BP describes the motion of a test particle in the frame co-rotating with the binary (see e.g.\ \citet{ssdbook}).  In our problem the test particle moves in the same plane as the binary and the position vector with respect to the binary's center of mass has coordinates $(x,y)$. 
The CR3BP has an integral of motion known as Jacobi constant \citep{ssdbook}
\begin{equation}
\label{cj}
C=x^2+y^2+2\, \left( \frac{1-\mu}{r_0}+\frac{\mu}{r_2} \right)-\dot{x}^2-\dot{y}^2 \ ,
\end{equation}
where
\begin{eqnarray}
r_0^2 &=& (x+\mu)^2+y^2 \ , \\
r_2^2 &=& (1-\mu-x)^2+y^2 \ .
\end{eqnarray}

Due to our choice of initial conditions we have $y(0)=0$  and $\dot{x}(0)=0$, hence we can visualize the orbits using surfaces of section i.e.\ plotting $(x,\dot{x})$ when $y=0$ and $\dot{y}\times\dot{y}(0)>0$.
Since the test particle has   $e_1=0$ at $t=0$ then 
\begin{equation}
\label{cj0}
C=x(0)^2+2\, \left( \frac{1-\mu}{x(0)+\mu}+\frac{\mu}{1-\mu-x(0)} \right)-\dot{y}(0)^2 \  ,
\end{equation}
with
\begin{eqnarray}
x(0) &=& a_1-\mu \\ 
\dot{y}(0) &=& \pm \sqrt{\frac{1-\mu}{a_1}}-a_1
\end{eqnarray}
where the $\pm$ sign in $\dot{y}(0)$ applies to prograde and retrograde orbits, respectively.

The zero velocity curves (ZVC) are obtained by solving Eq.~(\ref{cj}) with $v^2=\dot{x}^2+\dot{y}^2=0$. These ZVC provide boundaries on the test particle's motion since it can only occur in the region with $v^2\ge0$.   
In particular, the ZVC at the collinear Lagrange point $L_1$ is the limit curve for motion solely around primary or secondary.  The ZVC at the collinear Lagrange points $L_2$ and $L_3$ are the limit curves that prevent escape from  $L_2$ or $L_3$, respectively. 
Therefore, comparing the test particle's Jacobi constant with the values at $L_1$, $L_2$ and $L_3$ provide us important information regarding stability. If $C>C_1$ the test particle must remain in orbit around the primary.  If $C<C_1$ collision with  secondary or primary stars is possible. If $C<C_2$ or $C<C_3$  escapes are possible through L2 or L3, respectively. 

Lagrange points have $\dot{x}=\dot{y}=0$. The collinear Lagrange points have $y=0$ while  the coordinate $x$ can be obtained as series expansions in $\mu$ \citep{ssdbook}. The  Jacobi constant at $L_1$, $L_2$, $L_3$, including terms up to 2nd order in $\mu$ is
\begin{eqnarray}
C_1 & \approx & 3+3^{4/3}\,\mu^{2/3}-\frac{10}{3}\,\mu+\frac{1}{9}\,3^{2/3}\,\mu^{4/3} \nonumber \\ 
 && -\frac{52}{81}\,3^{1/3}\,\mu^{5/3}+\frac{62}{81}\,\mu^2 \\
C_2 & \approx & 3+3^{4/3}\,\mu^{2/3}-\frac{14}{3}\,\mu+\frac{1}{9}\,3^{2/3}\,\mu^{4/3} \nonumber \\
&& -\frac{56}{81}\,3^{1/3}\,\mu^{5/3}+\frac{98}{81}\,\mu^2 \\
C_3 & \approx & 3+\mu-\frac{1}{48}\,\mu^2 \ .
\end{eqnarray} 
 In the next section we will plot the initial conditions that have $C=C_1$, $C=C_2$ and $C=C_3$
(where $C$ is given by Eq.~(\ref{cj0})). We will see that, as expected, these curves separate the regions of different end states for the test particle.

\section{Numerical stability study}

We constructed grids of initial conditions in the plane $(\alpha,\mu)$ with an step $\Delta\mu=0.002$ and $\Delta\alpha=0.005$ AU. Each point in the grid was then numerically integrated over $50\sim$ kyr (around 12000 binary periods, depending on $\mu$) using a Burlisch-Stoer based N-body code (precision better than $10^{-12}$) using  astrocentric osculating variables. During the integrations we computed the averaged MEGNO chaos indicator  $ \langle Y \rangle$ (MEGNO is the acronym of Mean  Exponential  Growth  of  Nearby  Orbits)  \citep{Cincotta_2000}.
 We show these MEGNO maps in Fig.\ref{prograde}(a) and Fig.~\ref{retrograde}(a). 
 
The MEGNO chaos maps use a threshold that is chosen in order to avoid excluding stable orbits that did not converge to their theoretical value or those orbits that are weakly chaotic. Thus the color scale shows ``stable" orbits in blue up to $\langle Y \rangle \approx 2.0$ (a particular choice based on integration of individual orbits for very long times and due to the characteristics of this system). 

MEGNO is a fast chaos indicator that allows to distinguish rapidly between regular and chaotic orbits. Within the integration time, all the orbits identified as unstable (red) either collide with primary or secondary, or escape from the system as we can see in Fig.~\ref{prograde}(b\&c) and Fig.~\ref{retrograde}(b\&c). 
The thin region colored with light-blue in the transition between chaos/regularity is typically unstable within $\sim 10000$ to $15000$ binary periods.
We integrated the grids with less resolution for $2.5\times10^5$ binary periods and no additional signs of instability were observed.

\subsection{Prograde case}

\begin{figure}
  \centering
    \includegraphics[width=9.0cm]{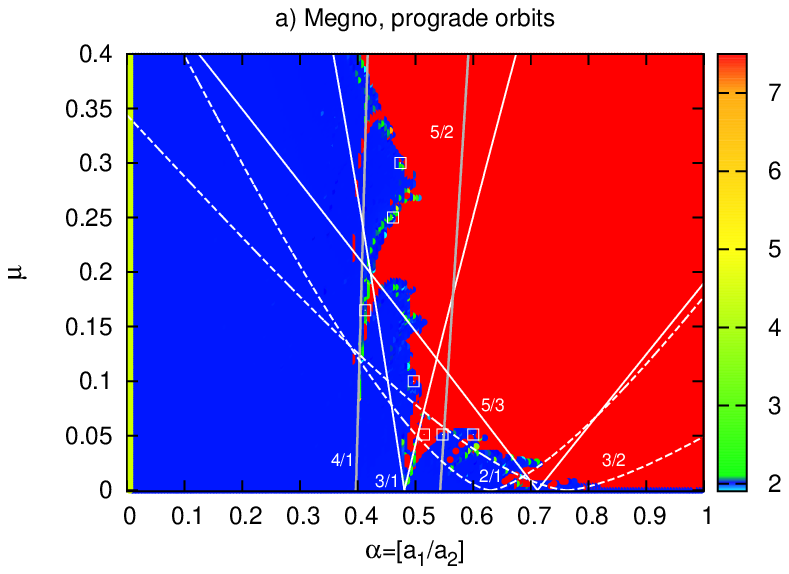} 
    \includegraphics[width=9.0cm]{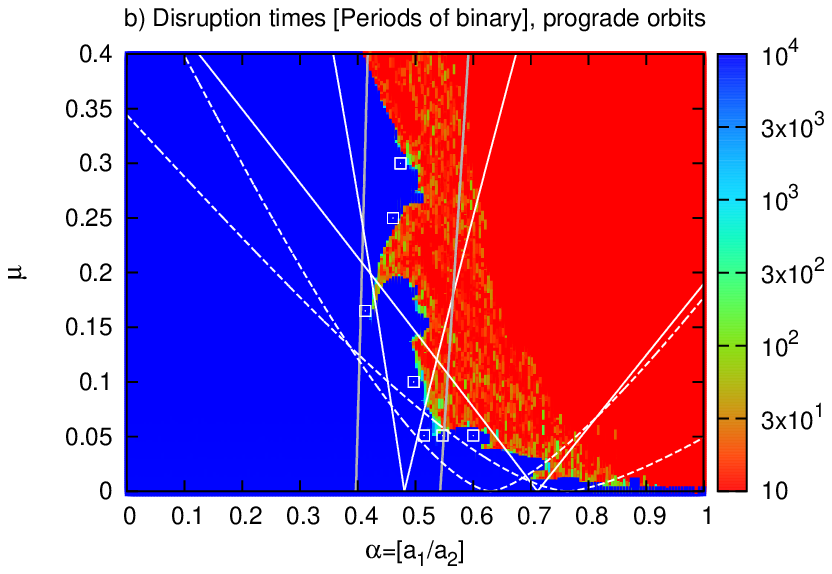} 
    \includegraphics[width=9.0cm]{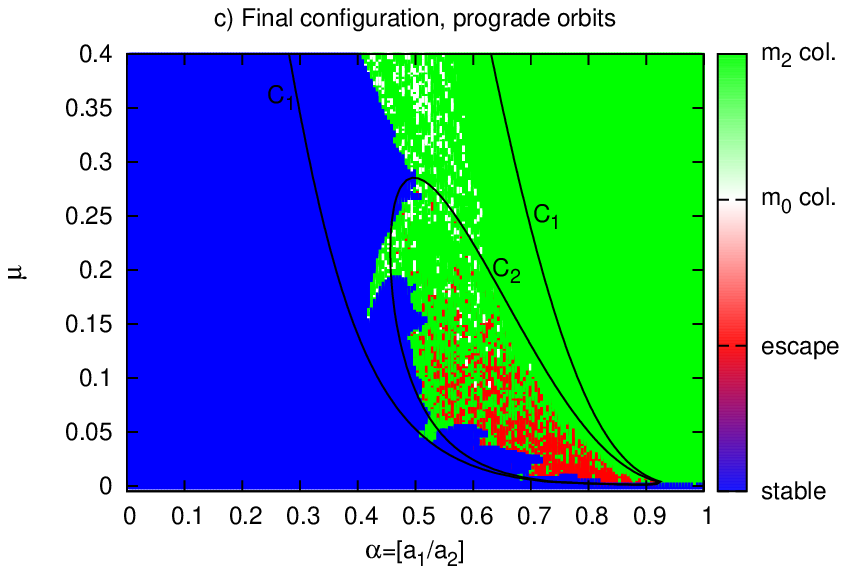} 
\caption{Dynamical analysis for prograde orbits. a) Stability  map  in  the  $(\alpha,\mu)$ space. Blue colors indicate stable orbits ($<Y> \leq 2.0$) while red colors indicate highly unstable orbits. b) Disruption times.
c) End state of the planet. 
The separatrices are shown in (a)\&(b) as dashed white lines (1st order resonances), solid white lines (2nd order resonances) and solid grey lines (3rd order resonances). The black solid curves in (c) indicate the initial conditions with $C=C_1$ and $C=C_2$.
The white squares at stability/instability transition zone indicate regions where we used the method of surfaces of section.
}    
\label{prograde}
\end{figure}

\begin{figure}
  \centering
    \includegraphics[width=9.0cm]{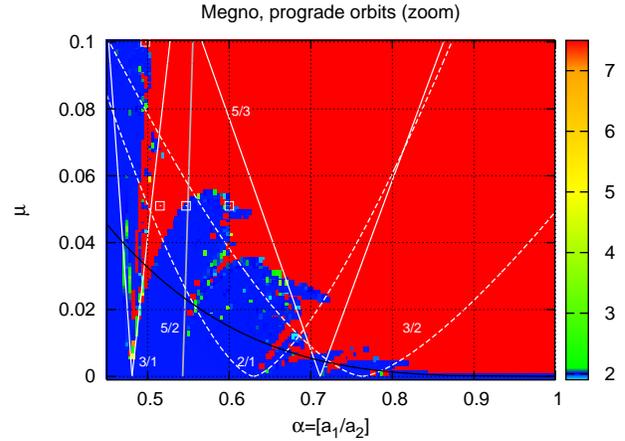}
\caption{Zoom of Fig.~\ref{prograde}(a). The separatrices are shown as dashed white lines (1st order resonances), solid white lines (2nd order resonances) and solid grey lines (3rd order resonances). The black solid curve $\alpha=1-1.33\,\mu^{2/7}$ is the 1st order resonance's overlap criterion \citep{Wisdom1980AJ}.}
\label{zoom}
\end{figure}

In Fig.~\ref{prograde} we show, for prograde orbits, the maps with: (a) MEGNO chaos indicator; (b) times of disruption of 3-body system; (c) planet end states (stable, collision or escape). In Fig.~\ref{zoom} we show a zoom of Fig.~\ref{prograde}(a).

 The separatrices of 1st order mean motion resonances (2/1 and 3/2)  are shown as white dashed lines in Figs.~\ref{prograde}(a)\&(b) and Fig.~\ref{zoom}.  The separatrices of 2nd order mean motion resonances (3/1 and  5/3) are shown as white solid lines in Figs.~\ref{prograde}(a)\&(b) and Fig.~\ref{zoom}. These separatrices are  obtained with Eq.~(\ref{wisdom}). The locations of 3rd order mean motion resonances (4/1 and 5/2), obtained by solving $\delta=0$  (Eq.~\ref{delta0}) for $\alpha$, are shown as solid grey lines in Figs.~\ref{prograde}(a)\&(b) and Fig.~\ref{zoom}. 
 
The initial conditions that have $C=C_1$ and $C=C_2$ are shown as black solid lines in Fig.~\ref{prograde}(c). The test particle's end states depend on the the Jacobi constant. Orbits with $C<C_1$ can escape through $L_1$ while orbits with $C<C_2$ can escape through $L_2$. Since $C>C_3$ escape through $L_3$ is not possible.
Moreover, the region near $\alpha=1$ has $C>C_1$ hence escape is not possible. However, initial conditions in this region correspond to orbits in a collision route with the secondary at $t=0$, thus they are unstable. 

From the perturbation theory, we predict that oscillations in $e_1$ at the 2/1 resonance are large enough for collision with the secondary when $\mu\ga 0.03$, since an initially circular orbit at exact resonance reaches $e_1$ such that $\alpha\,(1+e_1)\approx 1$ (where $e_{1}$ is obtained from Eq.~\ref{scaling00} with $R=2$, cf.~Fig.~\ref{h1}(a)).  This is in agreement with Figs.~\ref{prograde} \& \ref{zoom}.

We can see (Figs.~\ref{prograde} \& \ref{zoom}) that,  at small to moderate $\mu$ values, the border of the unstable region seems to coincide with resonances' locations, namely the 4/1 resonance at $\alpha\approx 0.4$ and the 3/1 resonance at $\alpha\approx 0.5$. We can also identify chaotic regions that seem to be associated with resonances' overlap (namely, between resonances 5/2 and 2/1, 2/1 and 5/3, and 5/3 and 3/2).  From Fig.~\ref{zoom} we see that the border of the unstable region when $\alpha\ga 0.7$  approximately coincides with the 1st order mean motion resonances' overlap criterion (black curve in Fig.~\ref{zoom}) in agreement with \citet{Wisdom1980AJ}. 
However, the perturbation method from which we obtained the resonance's locations and widths is certainly not valid when $\mu$ is large or when $\alpha\sim 1$.  Therefore, we will plot  surfaces of section, as described in Sect.~4, for initial conditions near the stability border (white squares in Fig.~\ref{prograde}).  The integration time for the surfaces of section is $10^5$ binary periods.

\begin{figure*}
 \begin{center}
   \begin{tabular}{l l}
    \includegraphics*[width=8cm]{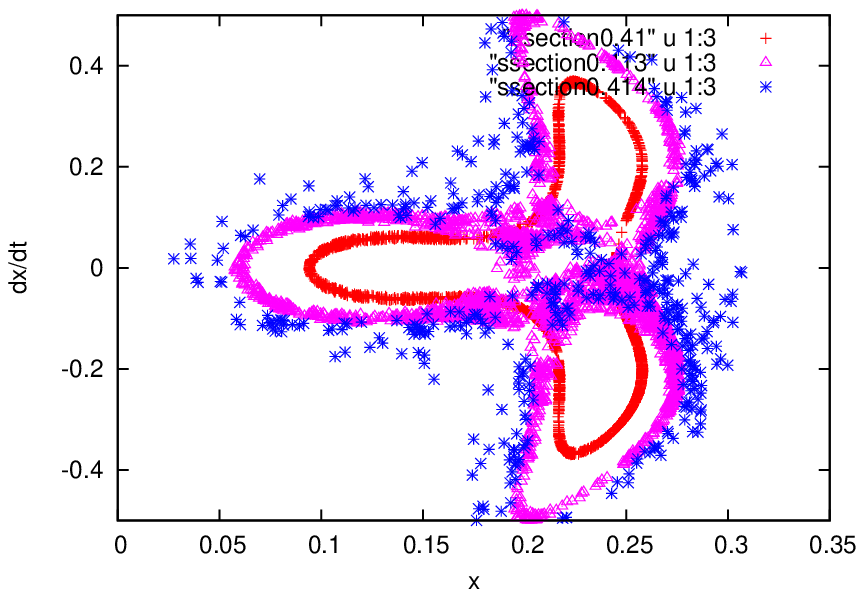} (a) & \includegraphics*[width=8cm]{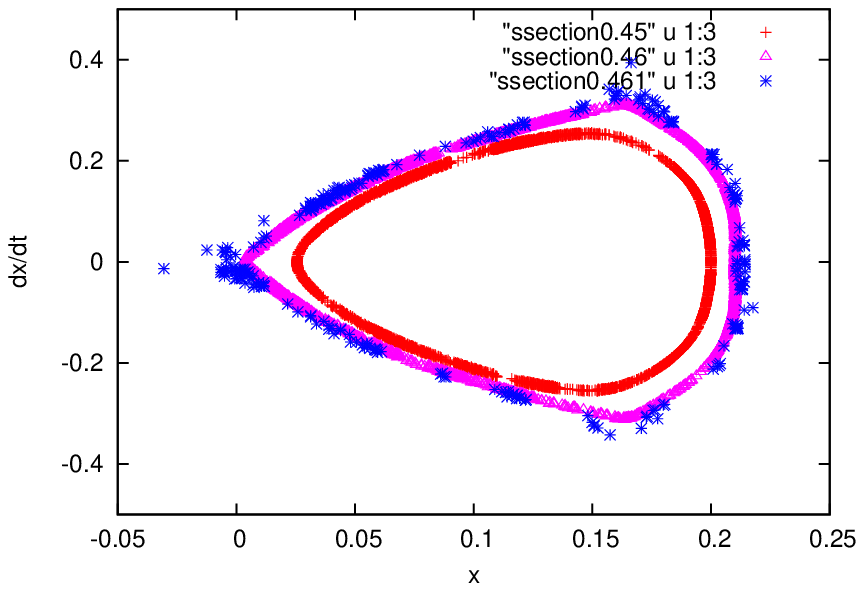} (b)  \\
    \includegraphics*[width=8cm]{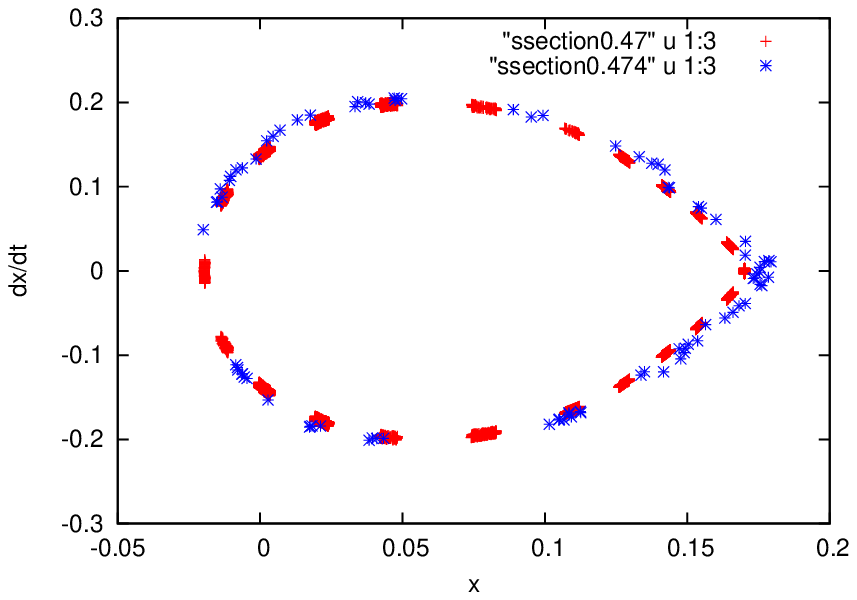} (c) &    \includegraphics*[width=8cm]{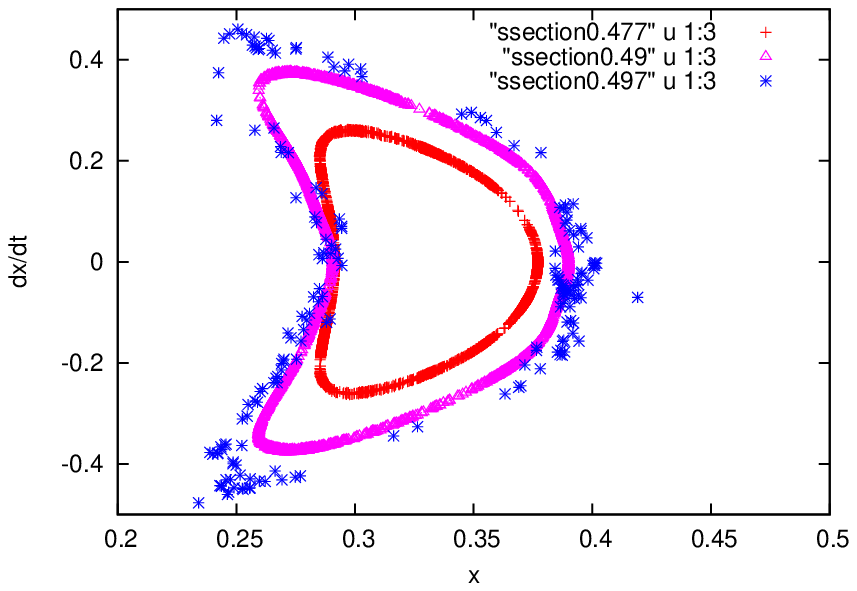}  (d) \\
    \includegraphics*[width=8cm]{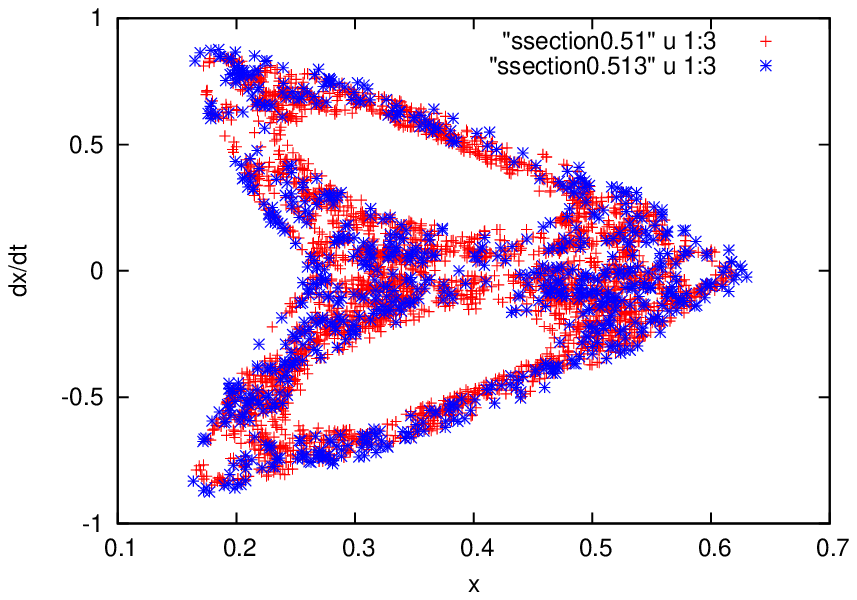} (e) &    \includegraphics*[width=8cm]{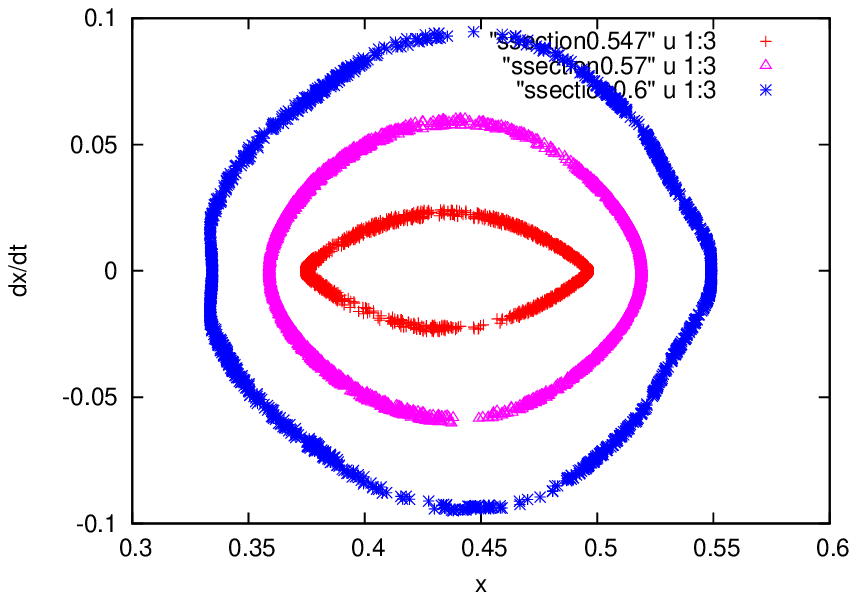} (f) \\
 \end{tabular}
\caption{Surfaces of section near stability boundary. 
(a) $\mu=0.165$: Orbit with  $\alpha=0.41$ (red) shows inner circulation in 4/1 resonance. Orbits with $\alpha=0.413$ (magenta) and $\alpha=0.414$ (blue) are chaotic due to 4/1 resonance separatrix crossing.  The blue orbit collides with the primary after 12740 binary periods. 
(b) $\mu=0.25$: Orbits with  $\alpha=0.45$ (red), $\alpha=0.46$ (magenta) and $\alpha=0.461$ (blue) are in the vicinity of the 4/1 resonance separatrix.  The blue orbit collides with the secondary after 8710 binary periods.
(c) $\mu=0.3$: Orbit with  $\alpha=0.47$ (red) is  librating in high order (22) resonance and nearby orbit with $\alpha=0.474$ (blue) is chaotic.  The blue orbit collides with the secondary after 1615 binary periods.
(d) $\mu=0.1$: Orbits with  $\alpha=0.477$ (red), $\alpha=0.49$ (magenta) and $\alpha=0.497$ (blue) are in vicinity of 3/1 resonance separatrix. The blue orbit collides with the secondary after 6800 binary periods. 
(e) $\mu=0.051$: Orbits with  $\alpha=0.51$ (red) and 
$\alpha=0.513$ (blue) are  in vicinity of 3/1 resonance separatrix.  The blue orbit collides with the secondary after 25100 binary periods.
(f) $\mu=0.051$: Orbits with  $\alpha=0.547$ (red), 
$\alpha=0.59$ (magenta) and $\alpha=0.60$ (blue). Stable orbits have $0.547 \le \alpha \le 0.6$. If $\alpha<0.547$ there is overlap of 5/2 and 2/1 resonances. If $\alpha>0.6$ eccentricity forcing at 2/1 resonance is  large enough for collision with secondary.
}
\label{fig1}
\end{center}
\end{figure*}

Comparing the initial conditions in Fig.~\ref{prograde} (white squares) with their follow up orbits in Fig.~\ref{fig1} we see that perturbation theory seems to be valid up to $\mu\approx0.2$ at the 4/1 resonance.  Following the discussion at the end of Sect.~3.1, we can extrapolate validity limits of $\mu\approx0.09$, $\mu\approx0.07$ and $\mu\approx0.03$  at the 3/1, 5/2 and 2/1 resonances, respectively.

In Fig.~\ref{fig1}(a) instability is associated with the 4/1 resonance separatrix ($\delta<0$ case; cf.~Fig.~\ref{h3}(c)).
In Fig.~\ref{fig1}(b) instability is also associated with the 4/1 resonance separatrix ($9>\delta>0$ case; cf.~Fig.~\ref{h3}(a)).
In Fig.~\ref{fig1}(c) we identify a high order (k=22) resonance.
In Figs.~\ref{fig1}(d)\&(e) instability is associated with the 3/1 resonance separatrix ($\delta<-4$ case; cf.~Fig.~\ref{h2}(c)).
In Fig.~\ref{fig1}(f) we show stable orbits associated with the 2/1 resonance ($0.547\le \alpha \le0.6$). When $\alpha<0.547$ there is chaos due to overlap with 5/2 resonance. When $\alpha>0.6$ eccentricity forcing at 2/1 resonance is large enough for collision with secondary, as seen above.

We conclude that instability for prograde orbits is either  due to single resonance forcing or resonance overlap. Regarding the latter mechanism, we infer resonance overlap from the observation of a main resonance's chaotic separatrix. We know from Chrikov's criterion that widespread chaos in Hamiltonian systems is caused by resonances overlapping. However, we cannot always identify the resonance(s) that overlap with the main resonance. These are likely to be high order resonances which are  difficult to identify in the surfaces of section, in particular in the chaotic regions where the resonances' overlap.

\begin{figure}
  \centering
   \includegraphics[width=8.0cm]{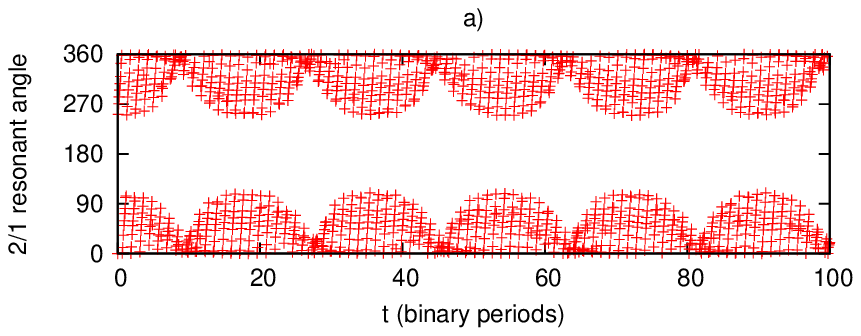} 
   \includegraphics[width=8.0cm]{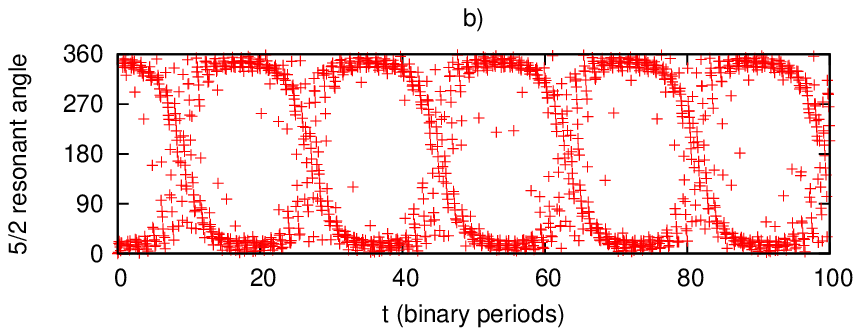}    
\caption{Proximity of 5/2 and 2/1 resonance overlap when $\mu=0.051$ and $\alpha=0.547$. Evolution of 2/1 resonance angle (a) and 5/2 resonance angle (b).}
\label{overlap} 
\end{figure}

In Fig.~\ref{overlap} we present a case where we identify the overlapping resonances. We show the evolution of the 2/1 and 5/2 resonant angles when  $\mu=0.051$ and $\alpha=0.547$ (red orbit in the surface of section Fig.~\ref{fig1}(f)). The 5/2 resonant angle alternates between libration and circulation since it is at the separatrix. When $\alpha<0.547$ both separatrices overlap and the orbits are unstable. These resonant angles are obtained from osculating elements with respect to the primary. When the perturbation from the secondary is small the osculating elements are approximately Keplerian in the short term.  Here, we can see that the resonant angles exhibit short term oscillations which indicates  that the  assumption of Keplerian osculating elements is not very good, despite the moderate mass ratio ($\mu=0.051$). Therefore, the osculating elements cannot be used to identify the resonances at larger mass ratio $\mu$ or when $\alpha\sim 1$. The method of surfaces of section is always valid thus it is  very useful to identify the main resonances and their chaotic separatrices.

\subsection{Retrograde case}

\begin{figure}
  \centering
    \includegraphics[width=9.0cm]{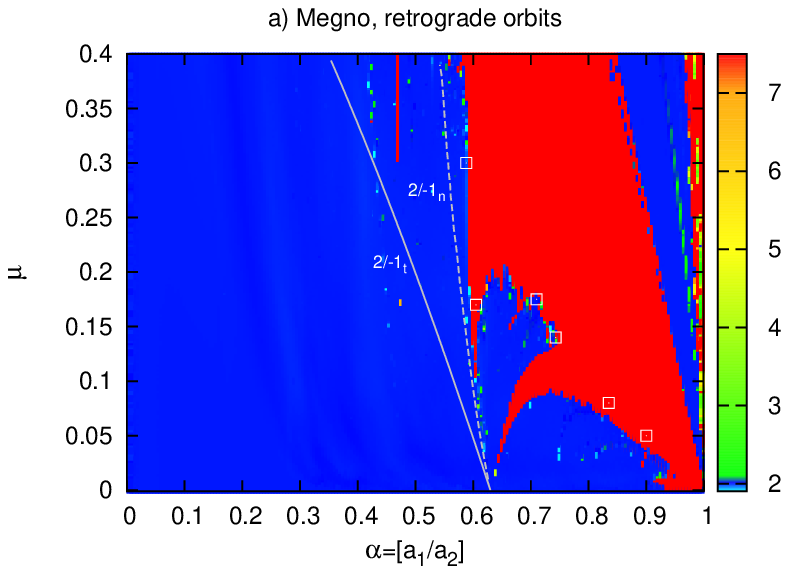} 
    \includegraphics[width=9.0cm]{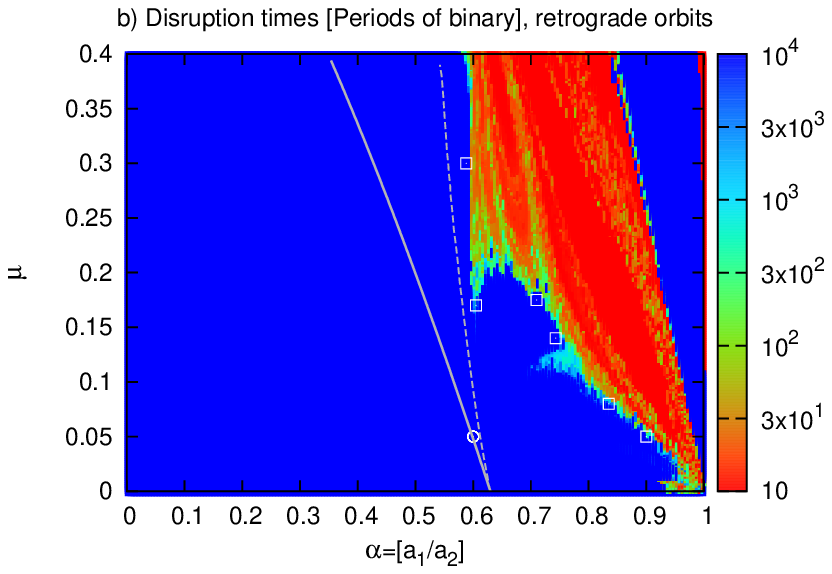} 
    \includegraphics[width=9.0cm]{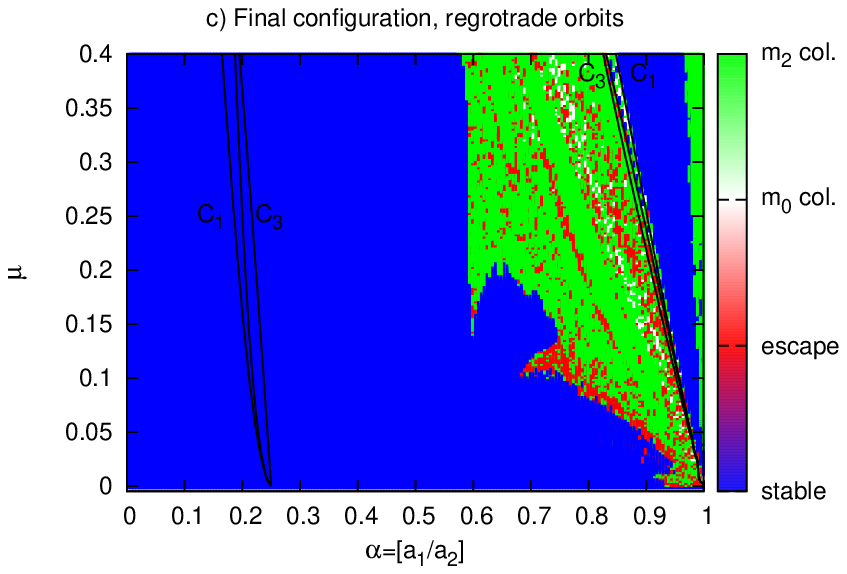}  
\caption{Dynamical analysis for retrograde orbits following same criteria as in Fig.~\ref{prograde}.
a) Stability  map  in  the  $(\alpha,\mu)$ space showing $<Y>$. 
b) Disruption times.
c) End state of the planet. 
 The theoretical 2/-1 resonance location shown in (a)\&(b) as white solid line is over-displaced to the left when $\mu\ga0.05$.  We show the ``corrected`` 2/-1 resonance location as white dashed line (see text for explanation). The black solid curves in (c) indicate the initial conditions with $C=C_1$, $C=C_2$ and $C=C_3$.
The white squares at stability/instability transition zone indicate regions where we used the method of surfaces of section.
The white circle is the initial condition for the red  orbit in Fig.~\ref{21res}.
}    
\label{retrograde}
\end{figure}

In Fig.~\ref{retrograde} we show, for retrograde orbits, the maps with: (a) MEGNO chaos indicator; (b) times of disruption of 3-body system; (c) planet end states (stable, collision or escape).

The location of the 2/-1 mean motion resonance, obtained by solving $\delta=0$  (Eq.~(\ref{delta0})) for $\alpha$ is shown as a white solid line  in Figs.~\ref{retrograde}(a)\&(b). For moderate to large $\mu$ values, Eq.~(\ref{precession}) over-estimates the precession rate hence it displaces the theoretical resonance location to the left. We obtain a ``corrected` 2/-1 mean motion resonance location  by solving $\delta=0$ for $\alpha$ while taking into account the precession rate measured in the numerical integrations (this is shown as a white dashed line in Figs.~\ref{retrograde}(a)\&(b)).

 The  initial conditions that have $C=C_1$, $C=C_2$ and $C=C_3$ are shown as black solid lines in Fig.~\ref{retrograde}(c) (these curves approximately coincide). We know that the test particle's end states depend on the the Jacobi constant. 
However, although orbits in between the curves $C=C_1$, $C=C_2$ or $C=C_3$ can escape through $L_1$, $L_2$ or $L_3$, respectively, in practice they only escape due to the effect of resonances.  The stable region near $\alpha=1$ corresponds to test particles orbiting the secondary at $t=0$, and is in agreement with the Jacobi constant criterion ($C>C_1$).

In Fig.~\ref{retrograde} we see that chaos and instability occurs at large values of the mass ratio $\mu$ or when 
$\alpha\sim 1$. However, from the discussion at the end of Sect.~3.1 and the results in Sect.~5.1, we conclude that perturbation theory is valid only up to $\mu\approx0.03$ at the 2/-1 resonance. This threshold is well below the instability region which at the 2/-1 resonance occurs only when $\mu>0.15$ (Fig.~\ref{retrograde}).   Therefore, perturbation theory cannot be used  and instead we will plot surfaces of section, as described in Sect.~4,  for initial conditions near the stability border (white squares in Fig.~\ref{retrograde}).  The integration time for the surfaces of section is $10^5$ binary periods.

\begin{figure*}
 \begin{center}
   \begin{tabular}{l l}
    \includegraphics*[width=8cm]{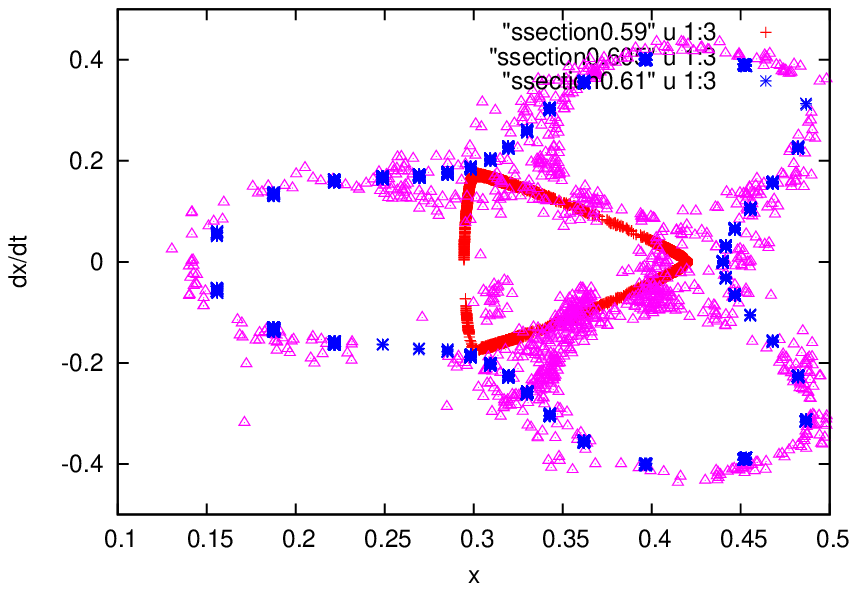} (a) & \includegraphics*[width=8cm]{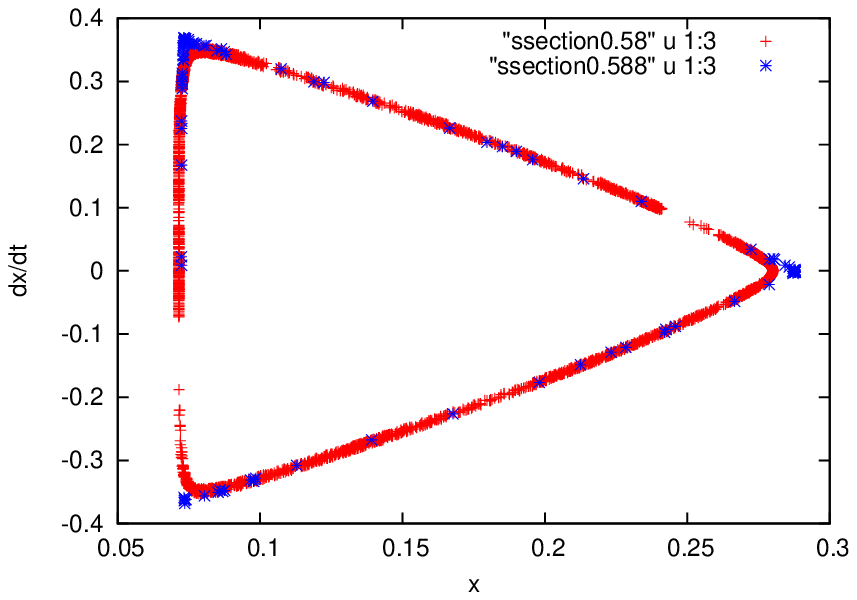} (b)  \\
    \includegraphics*[width=8cm]{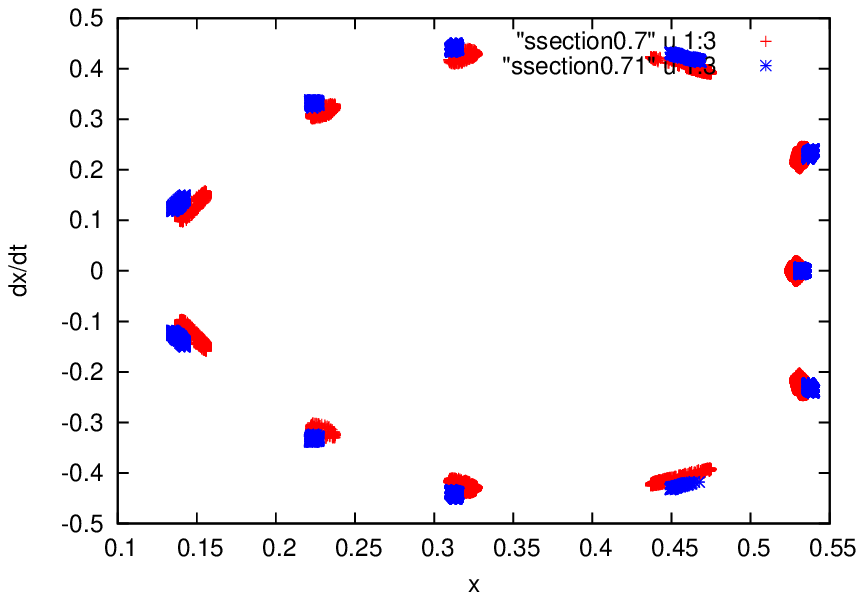} (c) &    \includegraphics*[width=8cm]{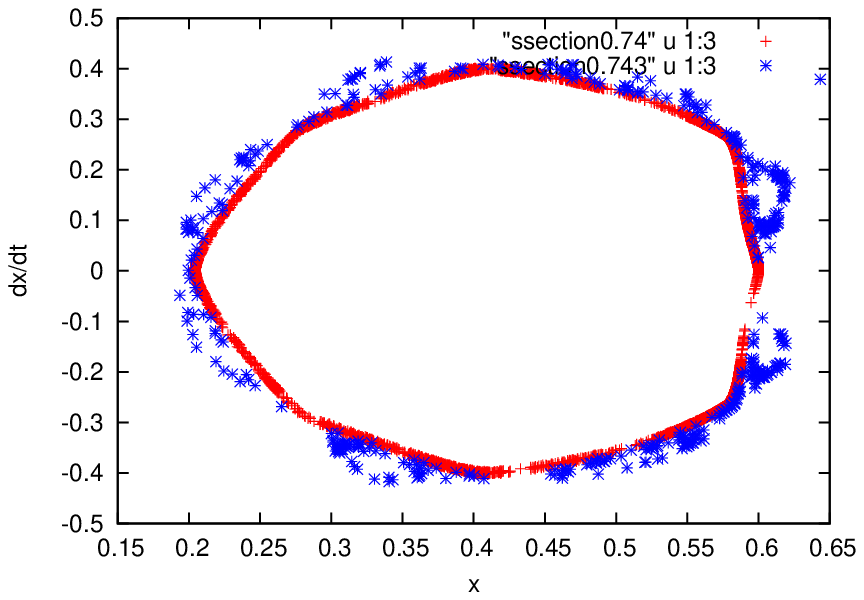} (d) \\
    \includegraphics*[width=8cm]{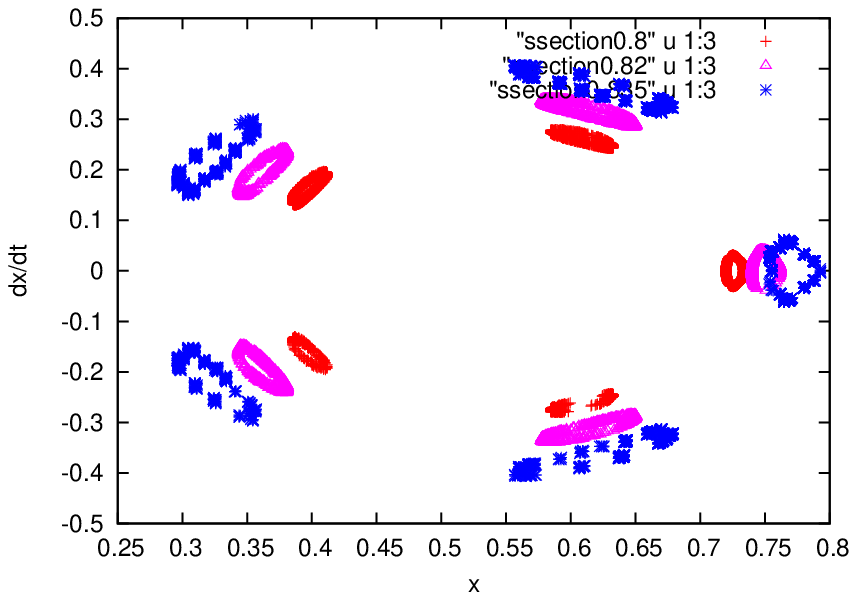} (e) &    \includegraphics*[width=8cm]{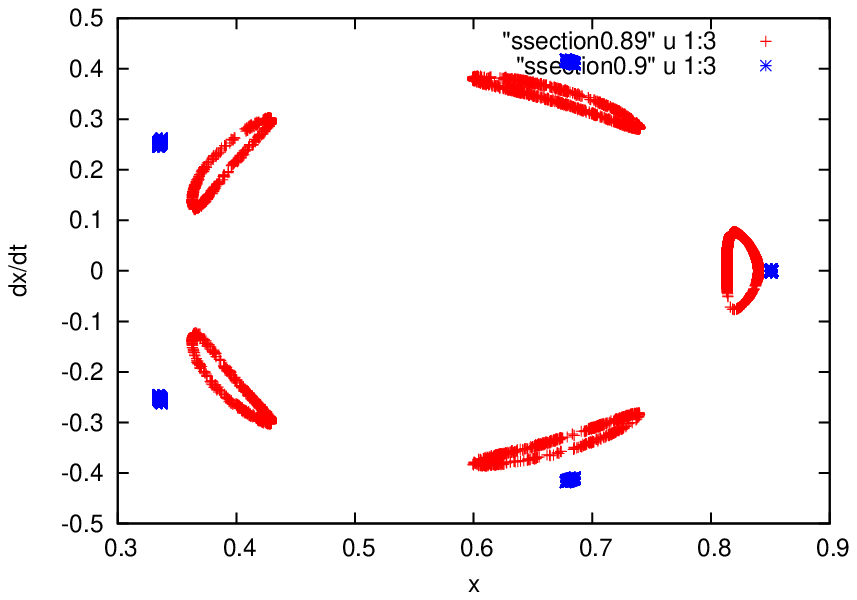} (f) \\
 \end{tabular}
\caption{Surfaces of section near stability boundary. 
(a) $\mu=0.17$: Orbits with  $\alpha=0.59$ (red), $\alpha=0.605$ (magenta) and $\alpha=0.61$ (blue) correspond, respectively, to inner circulation,  separatrix, and outer circulation of the 2/1 retrograde resonance. The magenta orbit collides with the secondary after 28500 binary periods.
(b) $\mu=0.3$: Orbit with  $\alpha=0.58$ (red) exhibits inner circulation in the 2/1 retrograde resonance. Orbit with $\alpha=0.588$ (blue) is at the separatrix of the 2/1 retrograde resonance and escapes after 600 binary periods.
(c) $\mu=0.175$: Orbit with  $\alpha=0.7$ (red) and $\alpha=0.71$ (blue) exhibit libration in 7/4 retrograde resonance.
(d) $\mu=0.14$:  Orbits with  $\alpha=0.74$ (red) and $\alpha=0.743$ (blue) correspond, respectively, to inner  circulation  and separatrix of 5/3 retrograde resonance. The blue orbit escapes after 8890 binary periods.
(e) $\mu=0.08$: Orbits with  $\alpha=0.8$ (red), $\alpha=0.82$ (magenta) and $\alpha=0.835$ (blue) exhibit  libration in 3/2 retrograde resonance.
(f) $\mu=0.05$: Orbits with $\alpha=0.89$ (red) and $\alpha=0.9$ (blue) exhibit libration in 3/2 retrograde resonance.
}
\label{fig1r}
\end{center}
\end{figure*}

In Figs.~\ref{fig1r}(a)\&(b) instability is associated with the 2/-1 resonance separatrix ($\delta<0$ case; cf.~Fig.~\ref{h3}(c)). As we noted above, in the retrograde case perturbation theory cannot be used to explain the instability border. In particular, the theoretical 2/-1 resonance location is over-displaced to the left as $\mu$ increases.  This is mostly due to Eq.~(\ref{precession}) over-estimating the precession rate. 
Taking this into account we obtained a ``corrected` resonance location  (white dashed line in Fig.~\ref{retrograde}(a)).

In Fig.~\ref{fig1r}(c)  the 7/-4 resonance causes oscillations in $e_1$ that lead to escape when $\alpha>0.71$.
In Fig.~\ref{fig1r}(d) instability is associated with the 5/-3 resonance separatrix. 
In Fig.~\ref{fig1r}(e)\&(f) the 3/-2 resonance causes oscillations in $e_1$ that lead to collision with the secondary when $\alpha>0.835$ and $\alpha>0.9$, respectively.

\begin{figure}
  \centering
    \includegraphics[width=8.0cm]{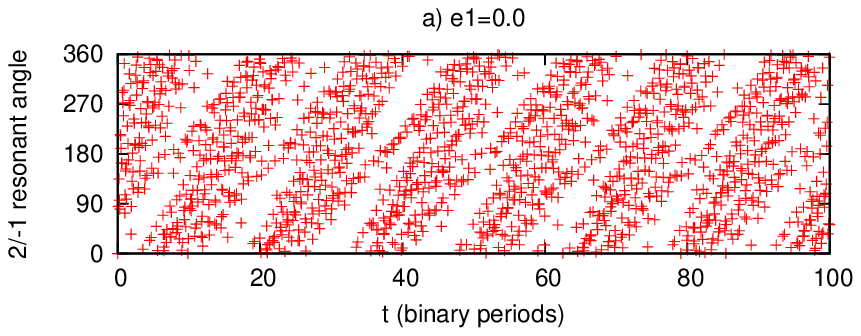} 
    \includegraphics[width=8.0cm]{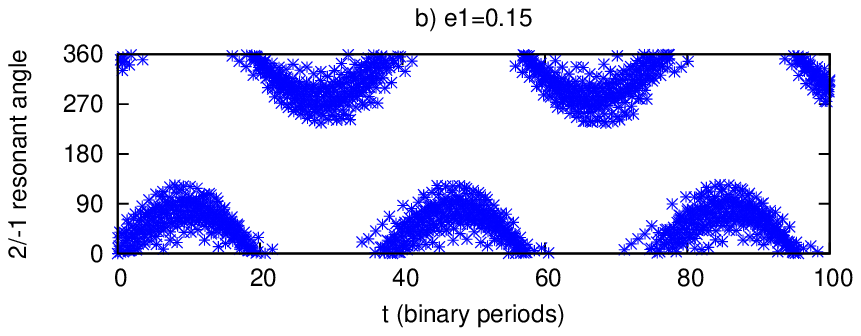} 
    \includegraphics[width=8.0cm]{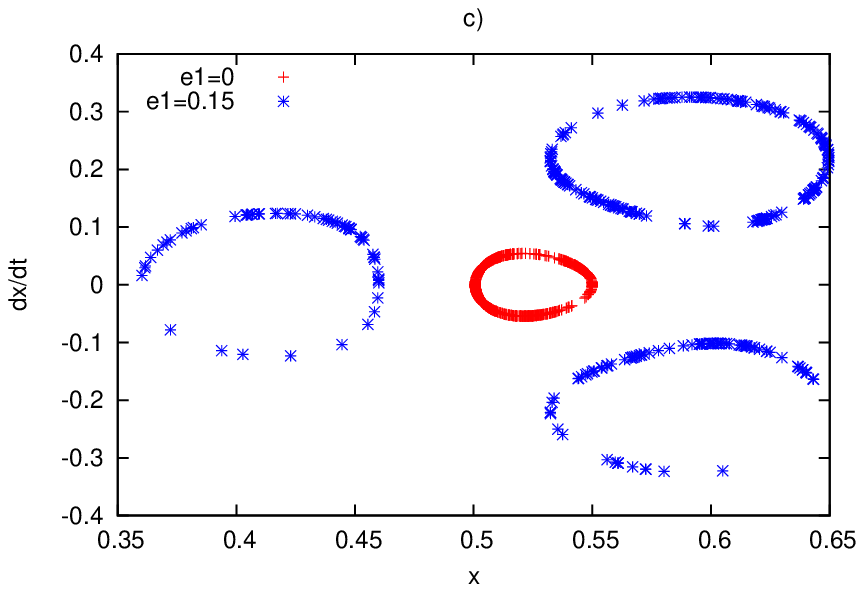} 
\caption{Evolution of 2/1 resonance angle when $\mu=0.05$ and $\alpha=0.6$: $e_1=0$ (a) and $e_1=0.15$ (b). Surface of section for both orbits (c).}
\label{21res}    
\end{figure}

In Fig.~\ref{21res} we show the evolution of the 2/-1 resonant angle when $\mu=0.05$  for orbits with $\alpha=0.6$, $e_1=0$ (a) and $e_1=0.15$ (b). When $e_1=0$ the resonant angle circulates and when $e_1=0.15$ the resonant angle librates.  The 2/-1 resonant angle also exhibits short period oscillations which indicate that the assumption of Keplerian osculating elements in the short term is not very good even at moderate values of the mass ratio ($\mu=0.05$). The surface of section (Fig~\ref{21res}(c)) confirms that these are regular orbits of the 2/-1 resonance, hence the spread of osculating elements is not due to chaotic diffusion.

\section{Discussion}

We investigated the stability of prograde and retrograde planets in circular binary star systems.
We saw that the cause of instability is  either increase of eccentricity due to single mean motion resonance forcing, or 
chaotic diffusion of eccentricity and semi-major axis due to overlap of adjacent mean motion resonances.

We computed the Jacobi constant in our grid of initial conditions and compared with the values at $L_1$, $L_2$, $L_3$. We saw that the boundaries of the instability regions are explained by single resonance forcing or by resonances' overlap. Nevertheless, in the prograde planet's simulations,  the  ZVC opens at $L_1$  near the instability border. However, in the retrograde planet's simulations, the ZVC opens at $L_1$, $L_2$ and $L_3$ when $\alpha\approx 0.2$ i.e.\ well before the instability border ($\alpha\approx 0.6$). We conclude that the ZVC opening at $L_1$, $L_2$ or $L_3$  are, as expected, necessary but not sufficient conditions for instability.  

\citet{Quarles_etal2011A&A} performed simulations of prograde orbits in binary systems and concluded that if the ZVC opens at $L_3$ then the orbit is unstable. This does not contradict our results since our initial conditions differ.  We start the planet at inferior conjuction  as viewed from the central star (i.e.\ between the 2 stars) while  \citet{Quarles_etal2011A&A}  start the planet at opposition. In particular, in our prograde simulations the ZVC never opens at $L_3$, while in our retrograde simulations a large set of orbits  with ZVC opening at $L_3$ are stable. 

Prograde planets are unstable from $\alpha\approx 0.4$ at $\mu>0.15$ and from $\alpha\approx 0.5$ at $0.15>\mu>0.05$.  The main causes of instability are:  overlap of the 4/1 resonance with nearby resonances at $\alpha\approx 0.4$ and $\mu>0.15$; overlap of the  3/1 resonance with nearby resonances at $\alpha\approx 0.5$ and $\mu>0.05$; overlap of the 5/2 and 2/1 resonances at $\alpha\approx 0.55$ and $\mu>0.05$; single resonance forcing in the 2/1 resonance at $\alpha\approx 0.6$; overlap of first order resonances when $\alpha\ga 0.7$ in agreement with \citet{Wisdom1980AJ}. 

Retrograde planets are unstable from $\alpha\approx 0.6$ and $\mu>0.15$ which coincides with the location of the 2/-1 resonance. The instability border at $\alpha\approx 0.7$ and $\mu=0.175$ coincides with the 7/-4 resonance.  The instability border at $\alpha\approx 0.74$ and $\mu=0.14$ coincides with the 5/-3 resonance separatrix. From $\alpha \approx 0.8$ instability is due to eccentricity forcing at the 3/-2  resonance.
 
\citet{Mudryk&Wu2006ApJ} identified the cause of prograde orbits' instability in eccentric binary systems as overlap of sub-resonances associated with certain mean motion ratios $p/q$. \citet{Mudryk&Wu2006ApJ}  also analyze previous low resolution numerical results of \citet{Holman&Wiegert1999AJ}  and argue that instability in circular binaries is due to overlap of sub-resonances  associated with the 3/1 resonance. However, this cannot work for circular binaries since there is only one resonant angle associated with a p/q resonance (all sub-resonances coincide).   Here, we identified the  cause of prograde orbits' instability in circular binary systems as single resonance forcing or overlap of different mean motion resonances, starting at the 4/1 resonance.  

We saw that retrograde planets are stable up to distances closer to the perturber than prograde planets.
We conclude that this is due to essential differences between the phase-space topology of retrograde versus prograde resonances. At mean motion ratio $p/q$, retrograde resonance has order $p+q$ while  prograde resonance has order $p-q$. Therefore, at a given resonance location $\alpha=(p/q)^{-2/3}$, we have: (a) eccentricity forcing on prograde planet is larger than eccentricity forcing on retrograde planet; (b) overlap with nearby resonances occurs at larger $\mu$ for retrograde configuration than prograde configuration.

\citet{Gayon&Bois2008} showed, using MEGNO, that retrograde resonance in 2 planet systems is more stable  than the equivalent prograde resonance. They conclude that this difference is due to close approaches being much faster and shorter for counter-revolving configurations than for the prograde ones \citep{Gayon&Bois2008}.  While this is true, we believe that the essential  difference is not the duration of close approaches but instead, as explained above, the phase-space topology of  prograde versus retrograde resonances.  An expansion of the Hamiltonian for retrograde resonance in 2 planet systems is presented in  \citet{Gayon_etal2009CMDA} but the numerical exploration of the model is limited to a small set of initial conditions and they do not conclude  on the essential differences between prograde and retrograde resonance. 

Finally, we refer that a similar mechanism could also explain the enhanced stability of retrograde satellites with respect to prograde satellites that has been observed, for instance, by \citet{Henon1970,Hamilton&Krivov1997,Nesvorny_etal2003AJ,Shen&Tremaine2008,Hinse_etal2010}.  However, satellite motion is a distinct problem from that presented in this article (planet within binary system) as the hierarchy of masses is very different.
 
\appendix

\section{}

Here, we follow partly the derivations in \citet{Peale1976}, \citet{Wisdom1980AJ}, \citet{Henrard&Lemaitre1983} and \citet{ssdbook} to model 1st, 2nd and 3rd order prograde resonances, and we extend this to model 3rd order retrograde resonance (2/-1).

The Hamiltonian of the CR3BP near $j/(j-k)$ prograde or retrograde resonance is
\begin{equation}
\label{h0}
H=-\frac{(1-\mu)}{2\,a_1}+U_{res}+U_{sec}
\end{equation}
where $G\,m_0=1-\mu=n_1^2\,a_1^3$ and $G\,m_2=\mu$.
The resonant term is
\begin{equation}
\label{resonant}
U_{res}= -\frac{\mu}{a_2}\,f_d(\alpha) e_1^k \cos((k-j)\,\lambda_1+j\,\lambda_2 - k\,\varpi_1 ) \ ,
\end{equation}
and the secular term is
\begin{equation}
\label{secular}
U_{sec}= -\frac{\mu}{a_2}\,f_s(\alpha) e_1^2 \ .
\end{equation}
From Lagrange's equations \citep{ssdbook}, $\varpi_1$ and $\lambda_1$ change due to the secular term, while $a_1$ and $e_1$ do not change; hence we can write
\begin{equation}
H=-\frac{(1-\mu)^2}{2\,\Lambda^2}+U_{res}-\Gamma\,\dot{\varpi}^{*}_{1}+\Lambda\,\dot{\lambda}^{*}_{1}
\end{equation}
where we used Poincar\'{e} canonical variables
\begin{eqnarray}
\Lambda &=& n_1\,a_1^2   \quad \quad \quad \quad \quad ,  \lambda_1 \\
\Gamma &=& n_1\,a_1^2  (1-\sqrt{1-e_1^2})  \quad , -\varpi_1  \ ,
\end{eqnarray}
and the secular variations in $\varpi_1$ and $\lambda_1$ are
\begin{eqnarray}
\label{sec1}
\dot{\varpi}^{*}_{1} &=&-\frac{\sqrt{1-e_1^2}}{n_1\,a_1^2\,e_1} \frac{\partial U_{sec}}{\partial e_1} \\
\label{sec2}
\dot{\lambda}^{*}_{1} & = &(1-\sqrt{1-e_1^2}) \dot{\varpi}^{*}_{1} \ .
\end{eqnarray}
 
Now,  we change to resonant variables 
\begin{eqnarray}
\Phi &=& \frac{k\,\Lambda-(k-j)\,\Gamma}{k} \quad , \phi=\lambda_1-\lambda_2\\
\Psi &=&  \Gamma \quad  , \psi= [ (k-j)\,\lambda_1+j\,\lambda_2 - k\,\varpi_1 ]/k 
\end{eqnarray}
via the generating function\footnote{For a mixed variable transformation: $\phi=\partial{F}/\partial{\Phi}$, $\psi=\partial{F}/\partial{\Psi}$, $\Lambda=\partial{F}/\partial{\lambda_1}$,  $\Gamma=-\partial{F}/\partial{\varpi_1}$.} 
\begin{equation}
F=\psi\,\Psi + \phi\,\Phi
\end{equation}
Since $F$  is time-dependent ($\lambda_2=\pm n_2\,t$) the transformation introduces the term $\partial{F}/\partial{t}$ in the new Hamiltonian, which is
\begin{eqnarray}
H &=& -\frac{(1-\mu)^2}{2\, \left( \Phi - \frac{(j-k)}{k} \Psi \right)^2}+U_{res} \pm \frac{j}{k}\,n_2\,\Psi \mp n_2\,\Phi \nonumber \\ 
&& - \Psi\,\dot{\varpi}^{*}_{1}+\left( \Phi - \frac{(j-k)}{k} \Psi \right)\,\dot{\lambda}^{*}_{1} \ .
\end{eqnarray}
Since $H$ does not depend on $\phi$, the momentum $\Phi$ is a conserved quantity.
Moreover,
if $e_1\ll 1$ then $\Psi \approx  \Phi\,e_1^2/2$ and thus we expand the 1st term in $H$  up to 2nd order around $\Psi=0$. Hence, dropping constant terms that depend only on $\Phi$, and changing the sign of $H$, we obtain\footnote{Where $(-1)^k$ is introduced because $\epsilon>0$ if $k$ even and $\epsilon<0$ if k odd.}

\begin{equation}
H = \gamma \Psi + \beta\,\Psi^2 +(-1)^k\,\epsilon (2\,\Psi)^{k/2} \cos(k\,\psi)
\end{equation}
where\footnote{Note that $\beta$ differs from expression in \citet{ssdbook} (no mass).}
\begin{eqnarray}
\gamma &=& [ (j-k)\, (n_1 +\dot{\lambda}^{*}_{1}) \mp j\,n_2 + k\,\dot{\varpi}^{*}_{1} ]/k \\
\beta &=&  \frac{3}{2}\frac{(j-k)^2}{k^2\,a_{1}^2} \\
\epsilon &=& \frac{\mu}{a_2}\,f_d(\alpha)\,n_{1}^{-k/2}\,a_{1}^{-k} \ .
\end{eqnarray}
As explained in \citet{ssdbook}, by introducing a scaled momentum
\begin{equation}
\label{scaling}
\bar{\Psi} = \left( \frac{\epsilon}{2\,\beta\,(-1)^k} \right)^{\frac{2}{k-4}}\Psi \ ,
\end{equation}
 this can be written as a single parameter Hamiltonian
\begin{equation}
\label{hamilton0}
H=\delta \bar{\Psi}+\bar{\Psi}^2+2\,(-1)^k\,(2\,\bar{\Psi})^{k/2}\,\cos(k\,\psi)
\end{equation}
with
\begin{equation}
\label{delta}
\delta = \gamma \left( \frac{4}{\epsilon^2\,\beta^{2-k}}\right)^{\frac{1}{4-k}} \ .
\end{equation}
The Hamiltonian (Eq.~\ref{hamilton0}) can be expressed in  cartesian canonical variables
$(x=R\,\cos(\psi),y=R\,\sin(\psi))$ where the scaling factor is
\begin{equation}
\label{scaling0}
R=\sqrt{2\,\bar{\Psi}} \ .
\end{equation}

\subsection*{Acknowledgements}
We thank Fathi Namouni for helpful discussions regarding the resonance Hamiltonian model. We thank the reviewer for helping to improve the clarity of the paper.
We acknowledge financial support from FCT-Portugal (grants PTDC/CTE-AST/098528/2008 and PEst-C/CTM/LA0025/2011).

\bibliographystyle{mn2e}

\bibliography{planetinbinary}

\end{document}